\documentclass[12pt,onecolumn]{IEEEtran}
\usepackage{graphicx,amssymb,amsmath}
\usepackage{multicol}
\usepackage[noadjust]{cite}
\usepackage{setspace}
\usepackage{stfloats}
\usepackage{amsthm}
\usepackage{flushend}
\usepackage{cases,subeqnarray}
\usepackage{bm,multirow,bigstrut}
\usepackage{textcomp}
\usepackage{latexsym,bm}
\usepackage{booktabs,changebar}
\usepackage{xcolor}
\usepackage{mathtools}
\usepackage{dsfont}
\usepackage{extarrows}
\usepackage{makecell}
\usepackage{booktabs}
\usepackage{multirow}
\usepackage{colortbl}
\definecolor{tabcolor}{rgb}{.105,.410,.113}
\makeatletter

\newcommand{\Rmnum}[1]{\expandafter\@slowromancap\romannumeral #1@}
\makeatother
\theoremstyle{plain}

\theoremstyle{plain}

\newtheorem{coro}{Corollary}

\IEEEoverridecommandlockouts

\begin{document}

%----------------------------title&author&tahnks----------------------------
\title{On the Performance of Dual-Hop Systems over Mixed FSO/mmWave Fading Channels}

\author{Yan Zhang, Jiayi~Zhang,~\IEEEmembership{Member,~IEEE}, Liang Yang,~\IEEEmembership{Member,~IEEE}, \\Bo Ai,~\IEEEmembership{Senior Member,~IEEE}, and Mohamed-Slim Alouini,~\IEEEmembership{Fellow,~IEEE}
\thanks{
		This work was supported in part by the National Key Research and Development Program under Grant 2016YFE0200900, in part by the Royal Society Newton Advanced Fellowship under Grant NA191006, in part by State Key Lab of Rail Traffic Control and Safety under Grants RCS2018ZZ007 and RCS2019ZZ007, in part by National Natural Science Foundation of China under Grants 61971027, U1834210, 61961130391, 61625106, and 61725101, in part by Beijing Natural Science Foundation under Grants 4182049 and L171005, and in part by the ZTE Corporation. (\textit{Corresponding Author: Jiayi Zhang.})}
\thanks{A conference version of this paper has been accepted in 2020 IEEE ICC workshop \cite{YZ2019fsommWave}.}
\thanks{Y. Zhang and J. Zhang are with the School of Electronic and Information Engineering, Beijing Jiaotong University, Beijing 100044, P. R. China. (e-mail: jiayizhang@bjtu.edu.cn).}
\thanks{L. Yang is with the College of Computer Science and Electronic Engineering, Hunan University, Changsha 410082, China (e-mail: liangyang.guangzhou@gmail.com).}
\thanks{B. Ai is with State Key Laboratory of Rail Traffic Control and Safety, Beijing Jiaotong University, Beijing 100044, China.(e-mail: boai@bjtu.edu.cn).}
\thanks{M.-S. Alouini is with the Computer, Electrical, and Mathematical Sciences and Engineering (CEMSE)
Division, King Abdullah University of Science and Technology (KAUST), Thuwal, Makkah Province, Saudi Arabia. (e-mail: slim.alouini@kaust.edu.sa)}
}

\maketitle

%----------------------------abstract----------------------------
\begin{abstract}
Free-space optical (FSO) links are considered as a cost-efficient way to fill the backhaul/fronthaul connectivity gap between millimeter wave (mmWave) access networks and optical fiber based central networks. In this paper, we investigate the end-to-end performance of dual-hop mixed FSO/mmWave systems to address this combined use. The FSO link is modeled as a Gamma-Gamma fading channel using both heterodyne detection and indirect modulation/direct detection with pointing error impairments, while the mmWave link experiences the fluctuating two-ray fading. Under the assumption of both amplify-and-forward and decode-and-forward relaying, we derive novel closed-form expressions for the outage probability, average bit error probability (BER), ergodic capacity, effective capacity in terms of bivariate Fox's $H$-functions. Additionally, we discuss the diversity gain and provide other important engineering insights based on the high signal-to-noise-ratio analysis of the outage probability and the average BER. Finally, all our analytical results are verified using Monte Carlo simulations.
\end{abstract}

%----------------------------keywords----------------------------

%\newpage
\IEEEpeerreviewmaketitle
\section{Introduction}
It is acknowledged that the dense deployment of small cells is one of the key architecture that enables the coverage of extremely high data rate in the fifth-generation (5G) wireless network. One significant concern in the deployment of such network is the backhauling which connects the massive data traffic from small cells to the core network. Free-space optical (FSO) has been considered as a feasible solution to the emerging backhaul/fronthaul requirements for ultra-dense heterogeneous small cells in 5G networks \cite{alzenad2018fso}. An FSO link can be deployed by setting a pair of laser-photodetector transceivers in line of sight between two points and it supports high data rate transmission. Due to its high security level at the unlicensed optical spectrum, FSO can deal with the issue of spectrum crunch in the backhaul link. In addition, the ability of immunity to electromagnetic interference makes FSO link a good solution to offer connectivity between radio frequency (RF) access network and optical fiber based central network. Heterodyne detection and intensity modulation/direct detection (IM/DD) are the two main modes of detection in FSO systems. In heterodyne detection, the received signal is mixed with a coherent signal of a laser beam produced by the local oscillator. The two beams fall on the photodetector by a beam splitter. The signal output from the photodetector contains a component with the difference-frequency between the coherent signal and the received signal, which is called heterodyne frequency. However, in IM/DD, the photodetector detects changes in the light intensity directly without using a local oscillator. Compared with IM/DD, heterodyne detection is more complex but significantly improves the sensitivity of photodetection \cite{hetedection}.
Moreover, it is important to mention that on the RF side, the millimeter wave (mmWave) technology is one of the most
important techniques for small cells in 5G cellular networks. It has a large spectrum to extend the network capacity massively. Therefore, FSO and RF technologies have been deployed together in the so-called mixed dual-hop FSO/RF systems to combine the advantages of RF access (low cost, flexible coverage) and FSO backhaul (high rate, low latency) \cite{chen2019novel}.

However, fluctuations in both phase and intensity of the received signals caused by atmospheric turbulence are major performance limiting factors in FSO communication \cite{2}. In addition, FSO communication is vulnerable to weather conditions, such as rain, aerosols, and particularly fog. Moreover, the pointing error caused by buildings sway phenomenon due to thermal expansion, dynamic wind load and weak earthquakes that all result in vibration of the transmitter beam and misalignment between transmitter and receiver may lead to a severe performance degradation over the FSO links. On the other side, RF links are limited by latency problems.
From another perspective, relaying technique which can be classified into amplify-and-forward (AF) and decode-and-forward (DF) relaying has been demonstrated as an efficient solution to increase the capacity for wireless communication systems as well as extending cost-efficient coverage. In DF relaying systems, the relay fully decodes the received signal and retransmits the decoded version into the second hop, while AF relays just amplify and forward the incoming signal without performing any sort of decoding, which is less complex in using \textcolor{black}{relays} \cite{1}.
\subsection{Related Work}
Considerable efforts have been made to study the end-to-end performance of dual-hop FSO/RF systems employing AF or DF relaying with both heterodyne detection and IM/DD \cite{NakaGamma1,NakaGamma2,NakaGamma3,GammaRay1,GammaRay2,DGGEGK,GGGeneralNaka,WeibullGamma}. In \cite{NakaGamma1}, \cite{NakaGamma2} and \cite{NakaGamma3}, the performance of an AF mixed FSO/RF relay network was analyzed over Gamma-Gamma and Nakagami-$m$ fading channels. Similar works have been done for mixed FSO/RF systems with AF relaying in \cite{GammaRay1} and \cite{GammaRay2} with the assumption that the FSO and RF links experience Gamma-Gamma and Rayleigh fading channels, respectively. The same system model over double generalized Gamma and extended generalized-$K$ fading models was studied in \cite{DGGEGK}. In \cite{GGGeneralNaka}, the authors studied the performance of a mixed FSO/RF system assuming Gamma-Gamma and Generalized Nakagami-$m$ fading models. Similarly, the performance of a mixed FSO/RF system over exponentiated-Weibull and Nakagami-$m$ fading channels was studied in \cite{WeibullGamma}.  However, these works failed to take mmWave into consideration in the RF link to fit the future wireless communications. Most recently, the performance of mixed FSO/mmWave systems has been investigated in \cite{GeneralGGNaka,5GRFFSO2017,mmWave2019}. In \cite{GeneralGGNaka}, the authors assumed M{\'a}laga-$\mathcal{M}$ fading models for the FSO links and Rician fading models for the mmWave RF links using AF relaying. Whereas \cite{5GRFFSO2017} proposed a mixed FSO/mmWave system without a relay assuming M{\'a}laga-$\mathcal{M}$ and Weibull for the FSO and mmWave RF links, respectively. In \cite{mmWave2019}, the authors investigated the performance of a mixed FSO/mmWave system with a M{\'a}laga-$\mathcal{M}$ distributed FSO channel and a generalized-$K$ distributed mmWave RF channel.

Although the results from \cite{GeneralGGNaka,5GRFFSO2017,mmWave2019} are insightful, these works fall short in modeling the random fluctuations suffered by the received signal accurately in mmWave RF links. Moreover, as a significant performance metric, the effective capacity of a mixed FSO/mmWave system has not been investigated in aforementioned literature. In this paper, we tackle the above issues by analyzing dual-hop mixed FSO/mmWave systems where the mmWave RF link experiences the fluctuating two-ray (FTR) fading and providing novel analysis of effective capacity. The FTR
fading model recently proposed in \cite{32} fits better with experimental mmWave channel modeling data than other conventional stochastic channel models. Moreover, FTR fading includes several well known fading distributions as either special or limiting cases including Rayleigh, Rician and Nakagami-$m$. Considering the FTR fading for the mmWave link, the mathematical challenge for deriving novel exact expressions of important performance metrics is hugely improved. Furthermore, we provide new insights for the considered system. To our best knowledge, the FTR fading has only been tested in [18] to fit the 28 GHz mmWave channel. However, due to the flexibility of FTR distribution, it is expected to be more useful than Rician and other fading for mmWave channels. This is a very important question which needs to be explored in the future.

\subsection{Contribution}
Motivated by these studies, we look into the performance of a dual-hop mixed FSO/mmWave system where the FSO and RF links experience Gamma-Gamma and FTR fading channels, respectively. The effect of pointing error on the FSO link is also taken into account. It is necessary to point out that compared to the conference paper \cite{YZ2019fsommWave}, which only focuses on the outage probability and average bit error rate of the mixed FSO/mmWave system, in this paper, we provide an extensive performance analysis framework of the mixed FSO/mmWave system, including novel exact closed-form expressions for ergodic capacity and effective capacity. These new results can provide useful insights to design practical dual-hop mixed FSO/mmWave communication systems. More specifically, ergodic capacity is widely employed to assess the maximum long-term achievable rate averaged over ergodic states of the time-varying fading channel \cite{ErgodicCapacity1}, \cite{ErgodicCapacity2}. In addition, the effective capacity takes into account the delay constraints imposed by emerging real-time applications, which have different quality of service requirements \cite{ergodic2019,eff2019}.

The major contributions of this paper can be summarized as follows: 1) employing AF and DF relaying, we derive new closed-form expressions for the outage probability, average BER, ergodic capacity and effective capacity of the considered system. 2) based on 1), the effects of the atmospheric turbulence, pointing errors, relaying techniques and fading figures on the mixed FSO/mmWave system performance are analyzed; 3) in order to get additional insights into the impact of system parameters, we present asymptotic expressions for the outage probability and the average BER at high signal-to-noise ratios (SNRs) to show the achievable diversity gain. Note that our derived results are general and can include the existing results \textcolor{black}{in the literature \cite{13}} as special cases, since our adopted RF fading channel in this paper is the most general one.

\subsection{Organization}
The remainder of this paper is organized as follows: In Section II, we introduce the system and channel models. In Section III, we obtain the end-to-end SNR statistics of AF and DF relaying. In Section IV, we derive closed-form expressions of outage probability, average BER, ergodic capacity, effective capacity followed by the asymptotic expressions at high SNRs. In Section V, some numerical and simulation results are presented to confirm the accuracy of derived expressions.  We finally conclude the paper in Section VI.

\section{System And Channel Models}

\begin{figure}[t]
\centering
\includegraphics[scale=0.8]{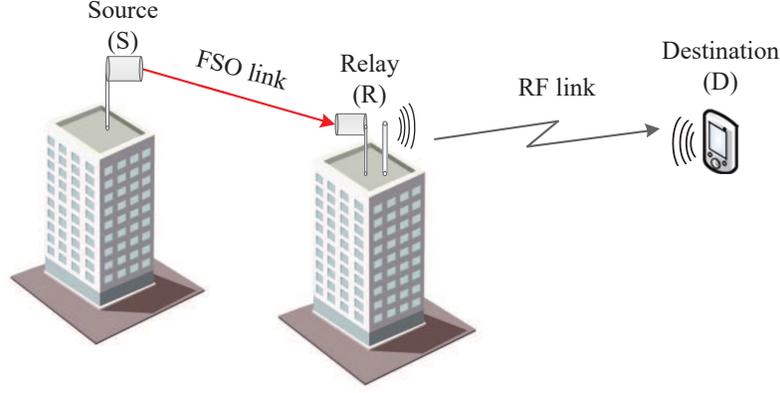}
\caption{A dual-hop system over mixed FSO/mmWave RF fading channels.}
\label{fig:systemmodel}
\end{figure}

We consider a dual-hop mixed FSO/mmWave communication system where the source node S communicates with the destination node D through an intermediate relay node R as illustrated in Fig. 1. The FSO link (S-R) is deployed for backhauling/fronthauling combined with the mmWave RF link (i.e., R-D link) for broadband radio access. Subcarrier intensity modulation (SIM) is employed in the source node to generate an optical signal. In the relay node, both heterodyne detection and IM/DD are considered to convert the received optical field to an electrical signal. The FSO link is assumed to follow a Gamma-Gamma fading distribution with pointing error. The probability density function (PDF) of the SNR, ${\gamma _{{\text{FSO}}}}$, is given by \cite[Eq. (3)]{zhang2015unified}
\begin{equation}
\label{FSO_PDF}
{f_{{\gamma _{\text{FSO}}}}}\!\left( \gamma  \right) \!=\! \frac{{{\xi ^2}\alpha \beta }}{{r\Gamma \!\left( \alpha  \!\right)\Gamma \!\left( \beta  \right)\gamma }}G_{1,3}^{3,0}\!\left(\! \!{\alpha \beta {{\left(\! {\frac{\gamma }{{{\mu _r}}}} \right)}^{\frac{1}{r}}}\left| \!\!{\begin{array}{*{20}{c}}
{{\xi ^2} + 1}\\
{{\xi ^2},\alpha ,\beta }
\end{array}} \right.} \!\!\!\right),
\end{equation}
where $\xi $ denotes the ratio between the equivalent beam radius at the receiver and the pointing error displacement standard deviation (jitter) at the receiver given as $\xi \!= \!\frac{{{\omega _{{z_{eq}}}}}}{{2{\sigma _s}}}$, with ${{\omega _{{z_{eq}}}}}$ and ${\sigma _s^2}$ represent the equivalent beam radius and the jitter variance at the receiver, respectively \cite{33}. Note that when $\xi\! \to \!\infty $, (1) converges to negligible pointing errors case. $r$ defines the mode of detection being used (i.e., $r\! =\! 1$ represents heterodyne detection and $r \!= \!2$ represents IM/DD) \cite{GGGeneralNaka}, $\Gamma \left( . \right)$ is the Gamma function as defined in \cite[Eq. (8.310)]{19}, $G\left( . \right)$ is the Meijer's $G$-function as defined in \cite[Eq. (9.301)]{19} and ${\mu _r}$ refers to the electrical SNR of the FSO link. Particularly, for $r \!= \!1$, ${\mu _1} \!=\! {\mu _{{\rm{heterodyne}}}} \!=\! {\bar \gamma _1}$, and for $r \!= \!2$, ${\mu _2} = {\mu _{{\rm{IM/DD}}}} = {{{{\bar \gamma }_1}\alpha \beta {\xi ^2}\left( {{\xi ^2} + 2} \right)} \mathord{\left/
 {\vphantom {{{{\bar \gamma }_1}\alpha \beta {\xi ^2}\left( {{\xi ^2} + 2} \right)} {\left[ {\left( {\alpha  + 1} \right)\!\left( {\beta  + 1} \right)\!{{\left( {{\xi ^2}{\rm{ + }}1} \right)}^2}} \right]}}} \right.
 \kern-\nulldelimiterspace} {\left[ {\left( {\alpha  + 1} \right)\left( {\beta  + 1} \right){{\left( {{\xi ^2}{\rm{ + }}1} \right)}^2}} \right]}}$, with the fading parameters $\alpha $ and $\beta $ related to the atmospheric turbulence conditions \cite{zhang2015ergodic}, and lower values of $\alpha $ and $\beta $ indicate severe atmospheric turbulence conditions. More specifically, when a plane wave propagation in the absence of inner scale is assumed, $\alpha $ and $\beta $ can be determined from the Rytov variance as $\alpha  \!=\! {\left[ {\exp \left( {\frac{{0.49\sigma _R^2}}{{{{\left( {1 + 1.11\sigma _R^{{{12} \mathord{\left/
 {\vphantom {{12} 5}} \right.
 \kern-\nulldelimiterspace} 5}}} \right)}^{{7 \mathord{\left/
 {\vphantom {7 6}} \right.
 \kern-\nulldelimiterspace} 6}}}}}} \right) - 1} \right]^{ - 1}}$ and $\beta  \!=\! {\left[ {\exp \left( {\frac{{0.51\sigma _R^2}}{{{{\left( {1 + 0.69\sigma _R^{{{12} \mathord{\left/
 {\vphantom {{12} 5}} \right.
 \kern-\nulldelimiterspace} 5}}} \right)}^{{5 \mathord{\left/
 {\vphantom {5 6}} \right.
 \kern-\nulldelimiterspace} 6}}}}}} \right) - 1} \right]^{ - 1}}$, where $\sigma _R^2 \!=\! 1.23{\left( {\frac{{2\pi }}{\lambda }} \right)^{{7 \mathord{\left/
 {\vphantom {7 6}} \right.
 \kern-\nulldelimiterspace} 6}}}C_n^2 L^{{{11} \mathord{\left/
 {\vphantom {{11} 6}} \right.
 \kern-\nulldelimiterspace} 6}}$ is the Rytov variance, $C_n^2$ is the refractive-index structure parameter, $\lambda $ is the wavelength, and $L$ represents the propagation distance  \cite{34}.
 By substituting (1) into ${F_{{\gamma _{\text{FSO}}}}}\left( \gamma  \right) = \int_0^\gamma  {{f_{{\gamma _{\text{FSO}}}}}\left( \gamma  \right)} d\gamma$ and utilizing \cite[Eq. (2.54)]{24}, the cumulative distribution function (CDF) of ${\gamma _{\text{FSO}}}$ can be written as
 \begin{align}
 \label{FSO_CDF}
&{F_{{\gamma _{\text{FSO}}}}}\left( \gamma  \right) = 1 - \frac{{{\xi ^2}}}{{\Gamma \left( \alpha  \right)\Gamma \left( \beta  \right)}} \nonumber\\
&\times H_{2,4}^{4,0}\left( \!{\frac{{{{\left( {\alpha \beta } \right)}^r}\gamma }}{{{\mu _r}}}\left| {\begin{array}{*{20}{c}}
{\left( {{\xi ^2} + 1,r} \right),\left( {1,1} \right)}\\
{\left( {0,1} \right),\left( {{\xi ^2},r} \right),\left( {\alpha ,r} \right),\left( {\beta ,r} \right)}
\end{array}} \right.}\!\! \right),
 \end{align}
where $H\left( . \right)$ is the Fox's $H$-function as defined in \cite[Eq. (1.1)]{24}.
It is assumed that the RF link experiences the FTR fading and the PDF of ${\gamma _{\text{RF}}}$ is given by \cite[Eq. 6]{28}
\begin{equation}
\label{RF_PDF_PRI}
{f_{{\gamma _{\text{RF}}}}}\left( \gamma  \right) = \frac{{{m^m}}}{{\Gamma \left( m \right)}}\sum\limits_{j = 0}^\infty  {\frac{{{K^j}{d_j}}}{{j!}}{f_G}\left( {\gamma ;j + 1,2{\sigma ^2}} \right)},
\end{equation}
where
\begin{equation}
\label{RF_PDF_1}
{f_G}\left( {\gamma ;j + 1,2{\sigma ^2}} \right) \buildrel \Delta \over = \frac{{{\gamma ^j}}}{{\Gamma \left( {j + 1} \right){{\left( {2{\sigma ^2}} \right)}^{j + 1}}}}\exp \left( { - \frac{\gamma }{{2{\sigma ^2}}}} \right),
\end{equation}
and
\begin{align}
%\label{RF_PDF_2}
&{d_j} \buildrel \Delta \over = \sum\limits_{k = 0}^j \!{\left(\!\!\! {\begin{array}{*{20}{c}}
j\\
k
\end{array}} \!\!\!\right)}{\left( \!{\frac{\Delta }{2}} \!\right)^k}\sum\limits_{l = 0}^k {\left(\!\!\! {\begin{array}{*{20}{c}}
k\\
l
\end{array}} \!\!\!\right)\Gamma \left( {j + m + 2l - k} \right){e^{\frac{{\pi \left( {2l - k} \right)i}}{2}}}} \nonumber\\
 &\times \!\!{\left(\!{{{\left( \!{m + K} \!\right)}^2}\! - \!{{\left( \!{K\Delta } \!\right)}^2}} \!\right)^{\frac{{ - \left( \!{j + m} \!\right)}}{2}}}\!\!P_{j + m - 1}^{k - 2l}\!\!\left( \!\!{\frac{{m + K}}{{\sqrt {{{\left(\! {m + K} \right)}^2} \!-\! {{\left( {K\Delta } \!\right)}^2}} }}} \!\!\right),\notag
\end{align}
where $P(.)$ denotes Legendre functions of the first kind \cite[Eq. (8.702)]{19}. Moreover, $K$ denotes the ratio of the average power of the dominant waves to the scattering multipath, $m$ is the fading severity parameter and $\Delta $ characterizes the similarity of two dominant waves varying from 0 to 1. In addition, the average SNR of RF link, ${\bar \gamma _{\text{RF}}}$, is defined as
\begin{align}
\label{FTR_average_SNR}
{\bar \gamma _{\text{RF}}} = \left( {{{{E_{\rm{b}}}} \mathord{\left/
 {\vphantom {{{E_{\rm{b}}}} {{N_0}}}} \right.
 \kern-\nulldelimiterspace} {{N_0}}}} \right)2{\sigma ^2}\left( {1 + K} \right),
\end{align}
 where ${{E_{\rm{b}}}}$ is the energy density.
Using \cite[Eq. (2.9.4), Eq. (2.1.5), and Eq. (2.1.4)]{27}, the PDF of the FTR distribution can be represented in terms of the Meijer's $G$-function as
\begin{equation}
\label{RF_PDF_MeijerG}
{f_{{\gamma _{\text{RF}}}}}\!\left( \gamma  \right) = \frac{{{m^m}}}{{\Gamma\! \left( m \right)}}\!\sum\limits_{j = 0}^\infty  {\frac{{{K^j}{d_j}}}{{j!\Gamma\! \left( {j + 1} \right)\gamma }}G_{0,1}^{1,0}\left[ {\frac{\gamma }{{2{\sigma ^2}}}\left|\! {\begin{array}{*{20}{c}}
 - \\
{j + 1}
\end{array}} \right.}\!\! \right]}.
\end{equation}
The CDF of the FTR distribution can be obtained by using \cite[Eq. (07.34.26.0008.01)]{Wolfram}, ${F_{{\gamma _{\text{RF}}}}}\left( \gamma  \right){\rm{ = }}\int_0^\gamma  {{f_{{\gamma _{\text{RF}}}}}\left( \gamma  \right)} d\gamma $, \cite[Eq. (2.54)]{24} and then utilizing \cite[Eq. (07.34.26.0008.01)]{Wolfram} again as
\begin{equation}
\label{RF_CDF}
{F_{{\gamma _{\text{RF}}}}}\!\left( \gamma  \right) = 1 \!-\! \frac{{{m^m}}}{{\Gamma \!\left( m \right)}}\!\sum\limits_{j = 0}^\infty  {\frac{{{K^j}{d_j}}}{{j!\Gamma \!\left( {j \!+\! 1} \right)}}} G_{1,2}^{2,0}\left[ {\frac{\gamma }{{2{\sigma ^2}}}\left| \!\!{\begin{array}{*{20}{c}}
1\\
{0,j \!+\! 1}
\end{array}} \right.} \!\!\!\right].
\end{equation}
Under the assumption of fixed-gain AF relaying, the end-to-end SNR can be written as \cite[Eq. 24]{DGGEGK}
\begin{equation}
\label{AF_END_TO_END_PRI}
{\gamma ^F} = \frac{{{\gamma _{\text{FSO}}}{\gamma _{\text{RF}}}}}{{{\gamma _{\text{RF}}} + {C_R}}},
\end{equation}
where ${C_R}$ represents a fixed relay gain.
The end-to-end SNR for DF relaying scenario can be derived as \cite[Eq. (26)]{DGGEGK}
\begin{equation}
\label{DF_END_TO_END_PRI}
{\gamma ^D} = \min \left( {{\gamma _{\text{FSO}}},{\gamma _{\text{RF}}}} \right).
\end{equation}

\section{End-To-End SNR Statistics}
\subsection{Fixed-Gain AF Relaying}
\subsubsection{Exact Result}
\begin{coro}\label{coro:CDF}
The CDF of the end-to-end SNR for a dual-hop mixed FSO/mmWave system using fixed-gain AF relay is
\begin{align}
\label{AF_End_to_End_CDF}
&{F_{{\gamma ^F}}}\left( \gamma  \right) = 1 - \frac{{{\xi ^2}{m^m}}}{{r\Gamma\! \left( \alpha  \right)\Gamma\! \left( \beta  \right)\Gamma \!\left( m \right)}}\sum\limits_{j = 0}^\infty  {\frac{{{K^j}{d_j}}}{{j!\Gamma \!\left( {j + 1} \right)}}} H_{1,0:0,2:3,2}^{0,1:2,0:0,3}\nonumber\\
& \times \left[\!\!\!\!\! {\left. {\begin{array}{*{20}{c}}
{\left( {1,1,\frac{1}{r}} \right)}\\
 - \\
{\begin{array}{*{20}{c}}
{\begin{array}{*{20}{c}}
 - \\
{\left( {0,1} \right)\left( {j + 1,1} \right)}
\end{array}}\\
{\left( {1 \!-\! {\xi ^2},1} \right)\left( {1 \!-\! \alpha ,1} \right)\left( {1 \!-\! \beta ,1} \right)}
\end{array}}\\
{\left( { - {\xi ^2},1} \right)\left( {0,\frac{1}{r}} \right)}
\end{array}} \!\!\right|\frac{{{C_R}}}{{2{\sigma ^2}}},\frac{1}{{\alpha \beta }}{{\left( {\frac{{{\mu _r}}}{\gamma }} \right)}^{\frac{1}{r}}}} \right].
\end{align}
\end{coro}
\begin{IEEEproof}
Please see Appendix A.
\end{IEEEproof}
Note that the Fox's $H$-function with two variables in \eqref{AF_End_to_End_CDF} can be calculated straightforwardly in well-known mathematical software, such as MATHEMATICA \cite[Eq. (1.1)]{23}. The MATLAB implementation of this function was provided in \cite{FOXH_MATLAB}. As a special case, for the RF link, when $\Delta  = 0$, $K \to \infty $, the FTR fading model reduces to Nakagami-$m$ fading model. Using \cite[Eq. (2.9.1)]{27} and setting $r = 1$, we can obtain the CDF of mixed Gamma-Gamma/Nakagami-$m$ systems using heterodyne detection with pointing errors as previous result \cite[Eq. (7)]{13}. In addition, by setting $\Delta  = 0$, $K \to \infty $, $r = 2$ and $\xi  \to \infty $, (10) simplifies to the special case where IM/DD is employed in the FSO link with no pointing errors and the RF link experiences Nakagami-$m$ fading.
Furthermore, by using \cite[Th. 1.7 and Th. 1.11]{27} along with \cite[Eqs. (1.5.9) and (1.8.4)]{27}, the CDF of the end-to-end SNR can be expressed in the asymptotic high-SNR regime after some algebraic manipulations as
\begin{equation}
\label{AF_End_to_End_CDF_ASY}
{F_{{\gamma ^F}}}\left( \gamma  \right) \approx \frac{{{\xi ^2}{m^m}}}{{\Gamma\! \left( \alpha  \right)\Gamma\! \left( \beta  \right)\Gamma\! \left( m \right)}}\!\sum\limits_{j = 0}^\infty  {\frac{{{K^j}{d_j}}}{{j!\Gamma \!\left( {j \!+\! 1} \right)}}} \!\sum\limits_{i = 1}^4 {{\psi _i}} \mu _r^{ - {\theta _i}},
\end{equation}
where ${\theta _i} = \left\{ {j + 1,\frac{{{\xi ^2}}}{r},\frac{\alpha }{r},\frac{\beta }{r}} \right\}$,
\begin{equation}
\label{AF_End_to_End_CDF_ASY1}
{\psi _1}  \buildrel \Delta \over =  \frac{{\Gamma \left[ {\alpha  - r\left( {j + 1} \right)} \right]\Gamma \left[ {\beta  - r\left( {j + 1} \right)} \right]}}{{\left( {j + 1} \right)\left[ {{\xi ^2} - r\left( {j + 1} \right)} \right]}}{\left( {\frac{{\gamma {C_R}{{\left( {\alpha \beta } \right)}^r}}}{{2{\sigma ^2}}}} \right)^{j + 1}},
\end{equation}
\begin{align}
\label{AF_End_to_End_CDF_ASY2}
{\psi _2} \buildrel \Delta \over =& \Gamma \left( {\alpha  - {\xi ^2}} \right)\Gamma \left( {\beta  - {\xi ^2}} \right){\left( {{\gamma ^{\frac{1}{r}}}\alpha \beta } \right)^{{\xi ^2}}}\nonumber\\
 &\times \left( {\frac{{\Gamma \left( {j + 1 - \frac{{{\xi ^2}}}{r}} \right)}}{{{\xi ^2}}}{{\left( {\frac{{{C_R}}}{{2{\sigma ^2}}}} \right)}^{\frac{{{\xi ^2}}}{r}}} + \frac{{\Gamma \left( {j + 1} \right)}}{{{\xi ^2}}}} \right),
\end{align}
\begin{align}
\label{AF_End_to_End_CDF_ASY3}
{\psi _3} \buildrel \Delta \over = &\frac{{\Gamma \left( {\beta  - \alpha } \right)}}{{{\xi ^2} - \alpha }}{\left( {{\gamma ^{\frac{1}{r}}}\alpha \beta } \right)^\alpha }\nonumber\\
 &\times \left( {\frac{{\Gamma \left( {j + 1 - \frac{\alpha }{r}} \right)}}{\alpha }{{\left( {\frac{{{C_R}}}{{2{\sigma ^2}}}} \right)}^{\frac{\alpha }{r}}} + \frac{{\Gamma \left( {j + 1} \right)}}{\alpha }} \right),
\end{align}
\begin{align}
\label{AF_End_to_End_CDF_ASY4}
{\psi _4} \buildrel \Delta \over =& \frac{{\Gamma \left( {\alpha  - \beta } \right)}}{{{\xi ^2} - \beta }}{\left( {{\gamma ^{\frac{1}{r}}}\alpha \beta } \right)^\beta }\nonumber\\
 &\times \left( {\frac{{\Gamma \left( {j + 1 - \frac{\beta }{r}} \right)}}{\beta }{{\left( {\frac{{{C_R}}}{{2{\sigma ^2}}}} \right)}^{\frac{\beta }{r}}} + \frac{{\Gamma \left( {j + 1} \right)}}{\beta }} \right).
\end{align}
\subsubsection{Truncation Error} By truncating \eqref{AF_End_to_End_CDF} up to the first $N$ terms, we have
\begin{align}
\label{AF_CDF_TRUNCATION_ERROR1}
&\mathop {{F_{{\gamma ^F}}}}\limits^ \wedge  \left( \gamma  \right) = 1 - \frac{{{\xi ^2}{m^m}}}{{r\Gamma \left( \alpha  \right)\Gamma \left( \beta  \right)\Gamma \left( m \right)}}\!\sum\limits_{j = 0}^N \! {\frac{{{K^j}{d_j}}}{{j!\Gamma \left( {j + 1} \right)}}} H_{1,0:0,2:3,2}^{0,1:2,0:0,3}\nonumber\\
 &\times \left[\!\!\! {\left. {\begin{array}{*{20}{c}}
{\left( {1,1,\frac{1}{r}} \right)}\\
 - \\
{\begin{array}{*{20}{c}}
{\begin{array}{*{20}{c}}
 - \\
{\left( {0,1} \right)\left( {j + 1,1} \right)}
\end{array}}\\
{\left( {1 - {\xi ^2},1} \right)\left( {1 - \alpha ,1} \right)\left( {1 - \beta ,1} \right)}
\end{array}}\\
{\left( { - {\xi ^2},1} \right)\left( {0,\frac{1}{r}} \right)}
\end{array}} \!\!\!\right|\frac{{{C_R}}}{{2{\sigma ^2}}},\frac{1}{{\alpha \beta }}{{\left( {\frac{{{\mu _r}}}{\gamma }} \right)}^{\frac{1}{r}}}} \right].
\end{align}
The truncation error of the area under the ${F_\gamma }\left( \gamma  \right)$ with respect to the first $N$ terms is given by
\begin{equation}
\label{AF_CDF_TRUNCATION_ERROR2}
\varepsilon \left( N \right) = {F_\gamma }\left( \infty  \right) - \mathop {{F_\gamma }}\limits^ \wedge  \left( \infty  \right).
\end{equation}
Table ${\rm \Rmnum{1}}$ shows the required terms $N$ for different channel parameters to demonstrate the convergence of the series in \eqref{AF_End_to_End_CDF}. For all considered cases, we \textcolor{black}{only need} less than 30 terms to achieve a satisfactory accuracy (e.g., smaller than ${10^{ - 3}}$).

\begin{table}[t]%图片位置，htbp分别代表here, top, bottom, page
  \label{TABLE_I}
  \centering%居中
  \caption{Required Terms $N$ For The Truncation Error $\left( {{\varepsilon } < {{10}^{ - 3}}} \right)$ With Different Parameters $K$, $m$ And $\Delta $}\label{table}%label是便于自己查看是那张表的，可不用，注释掉即可，表头一般放在表格上方
  \begin{tabular}{|c|c|c|}
  \hline
  \hline
  FTR fading parameters & $N$ & $\varepsilon $ \\
  \hline
  $K = 10,m = 2,\Delta  = 0.5$ & 18 & $9.6 \times {10^{ - 4}}$ \\
  \hline
  $K = 10,m = 0.3,\Delta  = 0.5$ & 23 & $9.5 \times {10^{ - 4}}$ \\
  \hline
 $K = 5,m = 8.5,\Delta  = 0.35$ & 14 & $2.6 \times {10^{ - 4}}$\\
 \hline
 \hline
\end{tabular}
\end{table}

\subsection{DF Relaying} Based on \eqref{DF_END_TO_END_PRI}, the CDF of the end-to-end SNR is given by \cite[Eq. (37)]{DGGEGK}
\begin{align}
\label{DF_End_to_End_CDF_PRI}
{F_{{\gamma ^D}}}\!\left( \gamma  \right) &= {F_{{\gamma _{\text{FSO}}}}}\left( \gamma  \right) + {F_{{\gamma _{\text{RF}}}}}\left( \gamma  \right) - {F_{{\gamma _{\text{FSO}}}}}\left( \gamma  \right){F_{{\gamma _{\text{RF}}}}}\left( \gamma  \right)\nonumber\\
 &= 1 - F_{{\gamma _{\text{FSO}}}}^C\left( \gamma  \right)F_{{\gamma _{\text{RF}}}}^C\left( \gamma  \right),
\end{align}
where $F_\gamma ^C\left( . \right)$ denotes complementary CDF (CCDF) of $\gamma$.
By substituting \eqref{FSO_CDF} and \eqref{RF_CDF} into \eqref{DF_End_to_End_CDF_PRI}, we obtain the CDF of dual-hop mixed FSO/mmWave systems employing DF relay as
\begin{align}
\label{DF_End_to_End_CDF}
&{F_{{\gamma ^D}}}\left( \gamma  \right) \!=\! 1\! - \!\frac{{{m^m}}}{{\Gamma \!\left( m \right)}}\!\sum\limits_{j = 0}^\infty \! {\frac{{{K^j}{d_j}}}{{j!\Gamma \!\left( {j \!+\! 1} \right)}}}\! H_{1,2}^{2,0}\!\left[ \!{\frac{\gamma }{{2{\sigma ^2}}}\!\!\left|\!\!\!{\begin{array}{*{20}{c}}
{\left( {1,1} \right)}\\
{\left( {0,1} \right),\left( {j \!+\! 1,1} \right)}
\end{array}} \!\!\!\!\right.} \right]\nonumber\\
 &\times \frac{{{\xi ^2}}}{{\Gamma \!\left( \alpha  \right)\Gamma\! \left( \beta  \right)}}H_{2,4}^{4,0}\left(\! {\frac{{{{\left(\! {\alpha \beta } \right)}^r}}}{{{\mu _r}}}\gamma \left| \!{\begin{array}{*{20}{c}}
{\left( {{\xi ^2} + 1,r} \right),\left( {1,1} \right)}\\
{\left( {0,1} \right),\left( {{\xi ^2},r} \right),\left( {\alpha ,r} \right),\left( {\beta ,r} \right)}
\end{array}} \!\right.} \!\!\!\right).
\end{align}
By using \cite[Eqs. (1.5.9) and (1.8.4)]{27} and after some algebraic manipulations, the CDF of the end-to-end SNR can be asymptotically expressed at high SNRs for DF relaying as shown by
\begin{equation}
\label{DF_End_to_End_CDF_ASY}
F_{{\gamma ^D}}^\infty \left( \gamma  \right)\! =\! F_{{\gamma _{\text{FSO}}}}^\infty \!\left( \gamma  \right) \!+\! F_{{\gamma _{\text{RF}}}}^\infty \!\left( \gamma  \right) \!-\! F_{{\gamma _{\text{FSO}}}}^\infty \!\left( \gamma  \right)F_{{\gamma _{\text{RF}}}}^\infty \!\left( \gamma  \right),
\end{equation}
where
\begin{align}
\label{DF_End_to_End_CDF_ASY1}
F_{{\gamma _{\text{FSO}}}}^\infty \left( \gamma  \right) &= \frac{{r\Gamma \left( {\alpha  - {\xi ^2}} \right)\Gamma \left( {\beta  - {\xi ^2}} \right)}}{{\Gamma \left( \alpha  \right)\Gamma \left( \beta  \right)}}{\left( {\frac{{{{\left( {\alpha \beta } \right)}^r}}}{{{\mu _r}}}\gamma } \right)^{\frac{{{\xi ^2}}}{r}}}\nonumber\\
 &+ \frac{{r{\xi ^2}\Gamma \left( {{\xi ^2} - \alpha } \right)\Gamma \left( {\beta  - \alpha } \right)}}{{\alpha \Gamma \!\left( \alpha  \right)\Gamma \!\left( \beta  \right)\Gamma \!\left( {{\xi ^2} + 1 - \alpha } \right)}}{\left(\! {\frac{{{{\left( {\alpha \beta } \right)}^r}}}{{{\mu _r}}}\gamma } \!\right)^{\frac{\alpha }{r}}}\nonumber\\
 &+ \frac{{r{\xi ^2}\Gamma \left( {{\xi ^2} - \beta } \right)\Gamma \left( {\alpha  - \beta } \right)}}{{\beta \Gamma \!\left( \alpha  \right)\Gamma \!\left( \beta  \right)\Gamma \!\left( {{\xi ^2} + 1 - \beta } \right)}}{\left(\! {\frac{{{{\left( {\alpha \beta } \right)}^r}}}{{{\mu _r}}}\gamma }\! \right)^{\frac{\beta }{r}}},
\end{align}
\begin{equation}
\label{DF_End_to_End_CDF_ASY2}
F_{{\gamma _{\text{RF}}}}^\infty \left( \gamma  \right) = \frac{{{m^m}}}{{\Gamma \left( m \right)}}\sum\limits_{j = 0}^\infty  {\frac{{{K^j}{d_j}}}{{j!\left( {j + 1} \right)\Gamma \left( {j + 1} \right)}}} {\left( {\frac{\gamma }{{2{\sigma ^2}}}} \right)^{j + 1}}.
\end{equation}
\textcolor{black}{By truncating \eqref{DF_End_to_End_CDF_ASY2} up to the first $N_1$ terms, we have}
{\color{black}\begin{align}
\mathop {F_{{\gamma _{\text{RF}}}}^\infty }\limits^ \wedge  \left( \gamma  \right) = \frac{{{m^m}}}{{\Gamma \left( m \right)}}\sum\limits_{j = 0}^{{N_1}} {\frac{{{K^j}{d_j}}}{{j!\left( {j + 1} \right)\Gamma \left( {j + 1} \right)}}} {\left( {\frac{\gamma }{{2{\sigma ^2}}}} \right)^{j + 1}}.
\end{align}}\textcolor{black}{The truncation error of the area under the $F_{{\gamma _{\text{RF}}}}^\infty \left( \gamma  \right)$ with respect to the first $N_1$ terms is given by}
{\color{black}\begin{align}
{\varepsilon _1}\left( {{N_1}} \right) = F_{{\gamma _{\text{RF}}}}^\infty \left( \infty  \right) - \mathop {F_{{\gamma _{\text{RF}}}}^\infty }\limits^ \wedge  \left( \infty  \right).
\end{align}}\textcolor{black}{In order to demonstrate the convergence of the infinite series in \eqref{DF_End_to_End_CDF_ASY2}, Table ${\rm \Rmnum{2}}$ presents the required truncation terms $N_1$ for different system and channel parameters. It should be noted that we only need less than 10 terms to converge the series for all considered cases and the truncation error is less than ${\rm{1}}{{\rm{0}}^{{\rm{ - 5}}}}$.}

\section{Performance Analysis}
\subsection{Fixed-Gain AF Relaying}
\subsubsection{Outage Probability} we encounter a situation labeled as outage when the instantaneous end-to-end SNR $\gamma $ falls below a given threshold ${\gamma _{\text{th}}}$, by replacing $\gamma $ with ${{\gamma _{\text{th}}}}$ in \eqref{AF_End_to_End_CDF}, we can easily obtain the outage probability as
\begin{equation}
\label{AF_OP}
P_{_{\text{out}}}^F\left( {{\gamma _{\text{th}}}} \right) = \Pr \left[ {{\gamma ^F} < {\gamma _{\text{th}}}} \right] = {F_{{\gamma ^F}}}\left( {{\gamma _{\text{th}}}} \right).
\end{equation}
\subsubsection{Average Bit-Error Rate} The average BER of a variety of binary schemes and non-binary modulation schemes can be expressed as \cite[Eq. (40)]{DGGEGK}
\begin{equation}
\label{AF_AEBR_PRI}
{{\bar P}_e} = \frac{\delta }{{2\Gamma \left( p \right)}}\sum\limits_{k = 1}^n {q_k^p\int_0^\infty  {{\gamma ^{p - 1}}\exp \left( { - {q_k}\gamma } \right)} } {F_\gamma }\left( \gamma  \right)d\gamma,
\end{equation}
where $\delta $, $p$, $n$ and ${{q_k}}$ denote different modulation schemes. For instance, $\left( {\delta ,p,{q_k},n} \right) = \left( {1,0.5,1,1} \right)$ denotes coherent binary phase
shift keying (CBPSK) and $\left( {\delta ,p,{q_k},n} \right) = \left( {1,1,1,1} \right)$  denotes differential BPSK (DBPSK).
%By Substituting \eqref{AF_End_to_End_CDF} into \eqref{AF_AEBR_PRI}, we have
\begin{coro}\label{coro:AEBR}
The average BER of the dual-hop mixed FSO/mmWave system is given by
\begin{align}
\label{AF_AEBR}
\bar P_e^F \!\!&=\! \!\frac{{n\delta }}{2} \!-\! \frac{{\delta {\xi ^2}{m^m}}}{{2r\Gamma \!\left( p \right)\!\Gamma\! \left( \alpha  \right)\!\Gamma\! \left( \beta  \right)\!\Gamma \!\left( m \right)}}\!\!\sum\limits_{j = 0}^\infty \! {\frac{{{K^j}{d_j}}}{{j!\Gamma \!\left( {j\! +\! 1} \right)}}} \!\!\sum\limits_{k = 1}^n\! {H_{1,0:0,2:3,3}^{0,1:2,0:1,3}} \nonumber\\
 &\!\times\! \left[\!\!\!\!\!\! \!{\left. {\begin{array}{*{20}{c}}
{\begin{array}{*{20}{c}}
{\begin{array}{*{20}{c}}
{\begin{array}{*{20}{c}}
{\left( {1,1,\frac{1}{r}} \right)}\\
 -
\end{array}}\\
 -
\end{array}}\\
{\left( {0,1} \right)\!\left( {j + 1,1} \right)}
\end{array}}\\
{\begin{array}{*{20}{c}}
{\left( {1 \!-\! {\xi ^2},1} \right)\!\left( {1 \!-\! \alpha ,1} \right)\!\left( {1 \!-\! \beta ,1} \right)}\\
{\left( {p,\frac{1}{r}} \right)\!\left( { - {\xi ^2},1} \right)\!\left( {0,\frac{1}{r}} \right)}
\end{array}}
\end{array}} \!\!\!\!\!\right|\frac{{{C_R}}}{{2{\sigma ^2}}},\frac{{{{\left( {{\mu _r}{q_k}} \right)}^{\frac{1}{r}}}}}{{\alpha \beta }}} \!\right].
\end{align}
\end{coro}
\begin{IEEEproof}
Please see Appendix B.
\end{IEEEproof}

\begin{table}[t]%图片位置，htbp分别代表here, top, bottom, page
  \arrayrulecolor{black}
  \centering%居中
  {\color{black}\caption{Required Terms ${N_1}$ For The Truncation Error (${\varepsilon _1} < {10^{ - 5}}$) With Different Parameters $K$, $m$ And $\Delta $}}\label{table2}%label是便于自己查看是那张表的，可不用，注释掉即可，表头一般放在表格上方
  \begin{tabular}{|c|c|c|}
  \hline
  \hline
  \textcolor{black}{FTR fading parameters} & \textcolor{black}{${N_1}$}& \textcolor{black}{$\varepsilon_1 $} \\
  \hline
  \textcolor{black}{$K = 10,m = 2,\Delta  = 0.5$} & \textcolor{black}{7} & \textcolor{black}{$3.1 \times {10^{ - 6}}$} \\
  \hline
  \textcolor{black}{$K = 10,m = 0.3,\Delta  = 0.5$} & \textcolor{black}{7} & \textcolor{black}{$1.2 \times {10^{ - 6}}$} \\
  \hline
 \textcolor{black}{$K = 5,m = 8.5,\Delta  = 0.35$} & \textcolor{black}{5} & \textcolor{black}{$8.1 \times {10^{ - 6}}$}\\
 \hline
 \hline
\end{tabular}
\end{table}

It is worth to mention that when we set $\Delta  = 0$, $K \to \infty $, $n = 1$, $\delta  = 1$ and $r = 1$, the BER in \eqref{AF_AEBR} simplifies to \cite[Eq. (13)]{13} where the FSO link is operating under heterodyne detection with pointing errors and the RF link experiences Nakagami-$m$ fading. Moreover, for $\Delta  = 0$, $K \to \infty $, $n = 1$, $\delta  = 1$, $r = 2$ and $\xi  \to \infty $, (25) reduces to the BER of a mixed Gamma-Gamma/Nakagami-$m$ system under IM/DD and no pointing errors. The asymptotic BER can be obtained by substituting \eqref{AF_End_to_End_CDF_ASY} into \eqref{AF_AEBR_PRI} and using \cite[Eq. (3.351.3)]{19} after some algebraic manipulations as
\begin{equation}
\label{AF_AEBR_ASY}
\bar P_e^F \!\approx\! \frac{{\delta {\xi ^2}{m^m}}}{{2\Gamma \!\left( p \right)\!\Gamma\! \left( \alpha  \right)\!\Gamma\! \left( \beta  \right)\!\Gamma \!\left( m \right)}}\!\sum\limits_{j = 0}^\infty \! {\frac{{{K^j}{d_j}}}{{j!\Gamma \!\left( {j \!+\! 1} \right)}}}\!{\sum\limits_{i = 1}^4\! {{\kappa _i}} \Gamma \!\left( {p\! +\! {\theta _i}} \right)},
\end{equation}
where
\begin{equation}
\label{AF_AEBR_ASY1}
{\kappa _1} \! \buildrel \Delta \over = \! \sum\limits_{k = 1}^n\! {\frac{{\Gamma \left[ {\alpha  - r\left( {j + 1} \right)} \right]\Gamma \left[ {\beta  - r\left( {j + 1} \right)} \right]}}{{\left( {j + 1} \right)\left[ {{\xi ^2} - r\left( {j + 1} \right)} \right]}}{{\left( {\frac{{{C_R}{{\left( {\alpha \beta } \right)}^r}}}{{2{\sigma ^2}{\mu _r}{q_k}}}} \right)}^{j + 1}}},
\end{equation}
\begin{align}
\label{AF_AEBR_ASY2}
{\kappa _2}\buildrel \Delta \over = &\sum\limits_{k = 1}^n {\Gamma \left( {\alpha  - {\xi ^2}} \right)\Gamma \left( {\beta  - {\xi ^2}} \right){{\left( {{{\left( {\frac{1}{{{\mu _r}{q_k}}}} \right)}^{\frac{1}{r}}}\alpha \beta } \right)}^{{\xi ^2}}}} \nonumber\\
 &\times \left( {\frac{{\Gamma \left( {j + 1 - \frac{{{\xi ^2}}}{r}} \right)}}{{{\xi ^2}}}{{\left( {\frac{{{C_R}}}{{2{\sigma ^2}}}} \right)}^{\frac{{{\xi ^2}}}{r}}} + \frac{{\Gamma \left( {j + 1} \right)}}{{{\xi ^2}}}} \right),
\end{align}
\begin{align}
\label{AF_AEBR_ASY3}
{\kappa _3} \buildrel \Delta \over = & \sum\limits_{k = 1}^n {\frac{{\Gamma \left( {\beta  - \alpha } \right)}}{{{\xi ^2} - \alpha }}{{\left( {{{\left( {\frac{1}{{{\mu _r}{q_k}}}} \right)}^{\frac{1}{r}}}\alpha \beta } \right)}^\alpha }} \nonumber\\
 &\times \left( {\frac{{\Gamma \left( {j + 1 - \frac{\alpha }{r}} \right)}}{\alpha }{{\left( {\frac{{{C_R}}}{{2{\sigma ^2}}}} \right)}^{\frac{\alpha }{r}}} + \frac{{\Gamma \left( {j + 1} \right)}}{\alpha }} \right),
\end{align}
\begin{align}
\label{AF_AEBR_ASY4}
{\kappa _4} = &\sum\limits_{k = 1}^n {\frac{{\Gamma \left( {\alpha  - \beta } \right)}}{{{\xi ^2} - \beta }}{{\left( {{{\left( {\frac{1}{{{\mu _r}{q_k}}}} \right)}^{\frac{1}{r}}}\alpha \beta } \right)}^\beta }} \nonumber\\
 &\times \left( {\frac{{\Gamma \left( {j + 1 - \frac{\beta }{r}} \right)}}{\beta }{{\left( {\frac{{{C_R}}}{{2{\sigma ^2}}}} \right)}^{\frac{\beta }{r}}} + \frac{{\Gamma \left( {j + 1} \right)}}{\beta }} \right).
\end{align}

\begin{table}[t]
 \arrayrulecolor{black}
  \centering
  {\color{black}\caption{Required Terms ${N_2}$ For The Truncation Error $\left( {{\varepsilon_2 } < {{10}^{ - 5}}} \right)$ With Different System And Channel Parameters}}\label{table3}
  \begin{tabular}{|c|c|c|}
  \hline
  \hline
  \textcolor{black}{System and channel parameters} & \textcolor{black}{$N_2$} & \textcolor{black}{$\varepsilon_2 $} \\
  \hline
  \textcolor{black}{\makecell[l]{$\alpha  = 5.42$,$\beta  = 3.8$,$\xi  = 5.0263$\\$K = 10$,$m = 2$,$\Delta  = 0.5$} }&\textcolor{black}{9} & \textcolor{black}{$1.1 \times {10^{ - 6}}$} \\
  \hline
  \textcolor{black}{\makecell[l]{$\alpha  = 3.446$, $\beta  = 1.032$, $\xi  = 5.0263$\\$K = 10$, $m = 2$, $\Delta  = 0.5$}} & \textcolor{black}{23} & \textcolor{black}{$9.9 \times {10^{ - 6}}$} \\
  \hline
 \textcolor{black}{\makecell[l]{$\alpha  = 5.42$, $\beta  = 3.8$, $\xi  = 0.893$\\$K = 10$, $m = 2$, $\Delta  = 0.5$}} & \textcolor{black}{27} &\textcolor{black}{$9.1 \times {10^{ - 6}}$}\\
 \hline
 \textcolor{black}{\makecell[l]{$\alpha  = 3.446$, $\beta  = 1.032$, $\xi  = 0.893$\\$K = 10$, $m = 2$, $\Delta  = 0.5$}} & \textcolor{black}{35} &\textcolor{black}{$8.8 \times {10^{ - 6}}$}\\
 \hline
 \textcolor{black}{\makecell[l]{$\alpha  = 5.42$, $\beta  = 3.8$, $\xi  = 0.893$\\$K = 10$, $m = 0.3$, $\Delta  = 0.5$}} & \textcolor{black}{21} & \textcolor{black}{$9.6 \times {10^{ - 6}}$}\\
 \hline
 \textcolor{black}{\makecell[l]{$\alpha  = 5.42$, $\beta  = 3.8$, $\xi  = 0.893$\\$K = 5$, $m = 8.5$, $\Delta  = 0.35$}} & \textcolor{black}{14} & \textcolor{black}{$8.2 \times {10^{ - 6}}$}\\
 \hline
 \hline
\end{tabular}
\end{table}
\textcolor{black}{By truncating \eqref{AF_AEBR_ASY} up to the first $N_2$ terms, we have}
{\color{black}\begin{align}
\mathop {\bar P_e^F}\limits^ \wedge  \!\! \approx \!\frac{{\delta {\xi ^2}{m^m}}}{{2\Gamma \!\left( p \right)\!\Gamma\! \left( \alpha  \right)\!\Gamma\! \left( \beta  \right)\!\Gamma\! \left( m \right)}}\!\!\sum\limits_{j = 0}^{{N_2}}\!  \frac{{{K^j}{d_j}}}{{j!\Gamma\! \left( {j + 1} \right)}}\!\!\sum\limits_{i = 1}^4 \! {\kappa _i}\Gamma \!\left( {p + {\theta _i}} \right).
\end{align}}\textcolor{black}{The truncation error of the area under the $\bar P_e^F$ with respect to the first $N_2$ terms is given by}
{\color{black}\begin{align}
{\varepsilon _2}\left( {{N_2}} \right) = \bar P_e^F - \mathop {\bar P_e^F}\limits^ \wedge.
\end{align}}\textcolor{black}{The required terms $N_2$ for different system and channel parameters are presented in Table ${\rm \Rmnum{3}}$ to demonstrate the convergence of the infinite series in \eqref{AF_AEBR_ASY}. We only need less than 40 terms to achieve a satisfactory accuracy (e.g., smaller than ${10^{ - 5}}$) for all considered cases.}
It can be seen from \eqref{AF_AEBR_ASY}-\eqref{AF_AEBR_ASY4} that the average BER decreases as the average SNR of both FSO (i.e., ${\mu _r}$) and RF (i.e., ${\bar \gamma _{\text{RF}}}$) links increase, which can be explained  from \eqref{FTR_average_SNR} that ${\bar \gamma _{\text{RF}}}$ is an increasing function of ${\sigma ^2}$ with $K$ fixed. Moreover, the performance of average BER degrades when the values of $\delta $ and $n$ get larger which represent for non-binary modulation schemes. Furthermore, it can be shown that the diversity gain is equal to
\begin{equation}
\label{AF_Diversity_Gain}
{G_d} = \min \left( {2,\frac{{{\xi ^2}}}{r},\frac{\alpha }{r},\frac{\beta }{r}} \right).
\end{equation}
We can observe from \eqref{AF_Diversity_Gain} that the diversity order is a function of FSO turbulence parameters (i.e., $\alpha $ and $\beta $), pointing error (i.e., $\xi $) and detection mode (i.e., $r$).

\subsubsection{Ergodic Capacity}The ergodic capacity is defined as $\bar C = \mathbb{E}\left[ {{{\log }_2}\left( {1 + c\gamma } \right)} \right]$, where $\mathbb{E}\left( . \right)$ refers to the expectation operator, $c = 1$ for heterodyne method (i.e., $r = 1$) and $c = {e \mathord{\left/
 {\vphantom {e {2\pi }}} \right.
 \kern-\nulldelimiterspace} {2\pi }}$
for IM/DD (i.e., $r = 2$).
By employing part-by-part integration method, ergodic capacity can be expressed in terms of the CCDF of $\gamma $ as
\begin{equation}
\label{AF_EC_PRI}
\bar C = \frac{c}{{\ln \left( 2 \right)}}\int_0^\infty  {\frac{{F_\gamma ^c\left( \gamma  \right)}}{{1 + c\gamma }}} d\gamma.
\end{equation}
The expression in \eqref{AF_EC_PRI} is exact for the case of heterodyne detection while it is a lower-bound for IM/DD since the transmitted symbols are always positive in IM/DD systems.
\begin{coro}\label{coro:capa}
The ergodic capacity of the dual-hop mixed FSO/mmWave systems can be derived as
\begin{align}
\label{AF_EC}
&\bar C^F = \frac{{{\xi ^2}{m^m}}}{{\ln \left( 2 \right)r\Gamma \left( \alpha  \right)\Gamma \left( \beta  \right)\Gamma \left( m \right)}}\sum\limits_{j = 0}^\infty  {\frac{{{K^j}{d_j}}}{{j!\Gamma \left( {j + 1} \right)}}} H_{1,0:0,2:4,3}^{0,1:2,0:1,4}\nonumber\\
 \!&\times\! \left[\!\!\!\!\!\!\! {\left. {\begin{array}{*{20}{c}}
{\begin{array}{*{20}{c}}
{\begin{array}{*{20}{c}}
{\begin{array}{*{20}{c}}
{\begin{array}{*{20}{c}}
{\left( {1,1,\frac{1}{r}} \right)}\\
 -
\end{array}}\\
 -
\end{array}}\\
{\left( {0,1} \right)\left( {j \!+\! 1,1} \right)}
\end{array}}\\
{\left( {1,\frac{1}{r}} \right)\left( {1 \!-\! {\xi ^2},1} \right)\left( {1 \!-\! \alpha ,1} \right)\left( {1 \!-\! \beta ,1} \right)}
\end{array}}\\
{\left( {1,\frac{1}{r}} \right)\left( { - {\xi ^2},1} \right)\left( {0,\frac{1}{r}} \right)}
\end{array}} \!\!\!\!\!\right|\frac{{{C_R}}}{{2{\sigma ^2}}},\frac{1}{{\alpha \beta }}{{\left( {{\mu _r}c} \right)}^{\frac{1}{r}}}} \!\right].
\end{align}
\end{coro}
\begin{IEEEproof}
Please see Appendix C.
\end{IEEEproof}

For $r = 1$, $\Delta  = 0$, $K \to \infty $, as a special case, \eqref{AF_EC} reduces to the ergodic capacity of a dual-hop mixed FSO/mmWave system where the RF link experiences Nakagami-$m$ fading under pointing error and heterodyne detection, given in \cite[Eq. (15)]{13}. We can also set $r = 2$, $\Delta  = 0$, $K \to \infty $ and $\xi  \to \infty $ to obtain the special case where the FSO link is under IM/DD for no pointing errors and the RF link experiences Nakagami-$m$ fading.

\newcounter{mytempeqncnt}
\begin{figure*}[b]
\normalsize
\vspace{-0.55cm}
\setcounter{mytempeqncnt}{\value{equation}}
\hrulefill
\vspace*{4pt}
\setcounter{equation}{44}
\begin{align}
\label{DF_EC}
&{{{\rm{\bar C}}}^D}{\rm{ = }}\frac{c}{{\ln \left( 2 \right)}}\frac{{{\xi ^2}{m^m}{\mu _r}}}{{\Gamma \left( \alpha  \right)\Gamma \left( \beta  \right)\Gamma \left( m \right){{\left( {\alpha \beta } \right)}^r}}}\sum\limits_{j = 0}^\infty  {\frac{{{K^j}{d_j}}}{{j!\Gamma \left( {j + 1} \right)}}} H_{4,2:1,1:1,2}^{0,4:1,1:2,0}\nonumber\\
 &\times \left(\!\!\! {\begin{array}{*{20}{c}}
{\left( {0;1,1} \right)\left( {1 - {\xi ^2} - r;r,r} \right)\left( {1 - \alpha  - r;r,r} \right)\left( {1 - \beta  - r;r,r} \right)}\\
{\left( { - {\xi ^2} - r;r,r} \right)\left( { - 1;1,1} \right)}
\end{array}\left| {\begin{array}{*{20}{c}}
{\left( {0,1} \right)}\\
{\left( {0,1} \right)}
\end{array}\left| {\begin{array}{*{20}{c}}
{\left( {1,1} \right)}\\
{\left( {0,1} \right)\left( {j + 1,1} \right)}
\end{array}\left| {\frac{{c{\mu _r}}}{{{{\left( {\alpha \beta } \right)}^r}}},\frac{{{\mu _r}}}{{2{\sigma ^2}{{\left( {\alpha \beta } \right)}^r}}}} \right.} \right.} \right.} \right).
\end{align}
\setcounter{equation}{\value{mytempeqncnt}}
\end{figure*}
\addtocounter{equation}{0}

\newcounter{mytempeqncnt2}
\begin{figure*}[b]
\normalsize
\vspace{-0.55cm}
\setcounter{mytempeqncnt}{\value{equation}}
\vspace*{4pt}
\setcounter{equation}{45}
\begin{align}
\label{DF_Effective}
&{R^D} =  - \frac{1}{A}{\log _2}\left( {1 - \frac{{{\xi ^2}{m^m}{\mu _r}}}{{\Gamma \left( \alpha  \right)\Gamma \left( \beta  \right)\Gamma \left( A \right)\Gamma \left( m \right){{\left( {\alpha \beta } \right)}^r}}}\sum\limits_{j = 0}^\infty  {\frac{{{K^j}{d_j}}}{{j!\Gamma \left( {j + 1} \right)}}} H_{4,2:1,1:1,2}^{0,4:1,1:2,0}} \right.\nonumber\\
&\left. { \times \left[ \!\!\!{\begin{array}{*{20}{c}}
{\left( {0;1,1} \right)\left( {1 - {\xi ^2} - r;r,r} \right)\left( {1 - \alpha  - r;r,r} \right)\left( {1 - \beta  - r;r,r} \right)}\\
{\left( { - {\xi ^2} - r;r,r} \right)\left( { - 1;1,1} \right)}
\end{array}\left| {\begin{array}{*{20}{c}}
{\left( { - A,1} \right)}\\
{\left( {0,1} \right)}
\end{array}\!\!\left| {\begin{array}{*{20}{c}}
{\left( {1,1} \right)}\\
{\left( {0,1} \right)\left( {j + 1,1} \right)}
\end{array}\!\!\left| {\frac{{{\mu _r}}}{{{{\left( {\alpha \beta } \right)}^r}}},\frac{{{\mu _r}}}{{2{\sigma ^2}{{\left( {\alpha \beta } \right)}^r}}}} \right.} \right.} \right.} \right]} \right).
\end{align}
\setcounter{equation}{\value{mytempeqncnt}}
\end{figure*}
\addtocounter{equation}{0}

\subsubsection{Effective Capacity}The effective capacity is defined as $R =  - \frac{1}{A}{\log _2}\left( {\mathbb{E}\left\{ {{{\left( {1 + \gamma } \right)}^{ - A}}} \right\}} \right)$, where $A \buildrel \Delta \over = {{\theta TB} \mathord{\left/
 {\vphantom {{\theta TB} {\ln 2}}} \right.
 \kern-\nulldelimiterspace} {\ln 2}}$ with the asymptotic decay rate of the buffer occupancy $\theta$, the block length $T$, and the system bandwidth $B$ \cite{zhang2015effective,7782422,8493599,7458872,7947095,6578148,8939132}. By employing part-by-part integration method, \textcolor{black}{the effective capacity} can be expressed in terms of the CCDF of $\gamma$ as
 \begin{equation}
\label{AF_EffectiveCapacity_PRI}
R =  - \frac{1}{A}{\log _2}\left( {1 - A\int_0^\infty  {{{\left( {1 + \gamma } \right)}^{ - A - 1}}} F_\gamma ^C\left( \gamma  \right)d\gamma } \right).
\end{equation}
\begin{coro}\label{coro:effective}
The effective capacity of the dual-hop mixed FSO/mmWave systems can be derived as
\begin{align}
\label{AF_EffectiveCapacity}
&{R^F} \!\!=\!\!  - \frac{1}{A}{\log _2}\!\!\left( \!\!{1 \!-\! \frac{{{\xi ^2}{m^m}}}{{r\Gamma\! \left( \alpha  \right)\!\Gamma\! \left( \beta  \right)\!\Gamma\! \left( m \right)\!\Gamma \!\left(A  \right)}}\!\!\sum\limits_{j = 0}^\infty \!\! {\frac{{{K^j}{d_j}}}{{j!\Gamma\! \left( {j \!\!+\!\! 1} \right)}}H_{1,0:0,2:4,3}^{0,1:2,0:1,4}} } \right.\nonumber\\
&\left. { \times \!\left[\!\!\!\!\!\!\!\!\!\! {\left. {\begin{array}{*{20}{c}}
{\left( {1,1,\frac{1}{r}} \right)}\\
{\begin{array}{*{20}{c}}
{\begin{array}{*{20}{c}}
{\begin{array}{*{20}{c}}
 - \\
 -
\end{array}}\\
{\left( {0,1} \right)\left( {j + 1,1} \right)}
\end{array}}\\
{\begin{array}{*{20}{c}}
{\left( {1 - A,\frac{1}{r}} \right)\left( {1 \!-\! {\xi ^2},1} \right)\left( {1 \!-\! \alpha ,1} \right)\left( {1 \!-\! \beta ,1} \right)}\\
{\left( {1,\frac{1}{r}} \right)\left( { - {\xi ^2},1} \right)\left( {0,\frac{1}{r}} \right)}
\end{array}}
\end{array}}
\end{array}}\!\!\!\!\!\!\!\!\! \right|\frac{{{C_R}}}{{2{\sigma ^2}}},\frac{{{{\left( {{\mu _r}} \right)}^{\frac{1}{r}}}}}{{\alpha \beta }}} \right]} \right).
\end{align}
\end{coro}
\begin{IEEEproof}
Please see Appendix D.
\end{IEEEproof}

It can be shown that when we set $\Delta  = 0$, $K \to \infty $, the effective capacity in \eqref{AF_EffectiveCapacity} can be simplified to the special case for Gamma-Gamma/Nakagami-$m$ fading channels using heterodyne detection which is given as
\begin{align}
\label{AF_EFF_Special_case}
&{R^F} =  - \frac{1}{A}{\log _2}\left( {1 - \frac{{{\xi ^2}}}{{\Gamma \left( \alpha  \right)\Gamma \left( \beta  \right)\Gamma \left( m \right)\Gamma \left( A \right)}}H_{1,0:0,2:4,3}^{0,1:2,0:1,4}} \right.\nonumber\\
&\left. { \times \left[ \!\!\!\!\!\!\!\!\!\!\!{\left. {\begin{array}{*{20}{c}}
{\left( {1,1,1} \right)}\\
{\begin{array}{*{20}{c}}
{\begin{array}{*{20}{c}}
{\begin{array}{*{20}{c}}
 - \\
 -
\end{array}}\\
{\left( {0,1} \right)\left( {m,1} \right)}
\end{array}}\\
{\begin{array}{*{20}{c}}
{\left( {1 - A,1} \right)\left( {1 - {\xi ^2},1} \right)\left( {1 - \alpha ,1} \right)\left( {1 - \beta ,1} \right)}\\
{\left( {1,1} \right)\left( { - {\xi ^2},1} \right)\left( {0,1} \right)}
\end{array}}
\end{array}}
\end{array}} \!\!\!\!\!\!\!\!\right|\frac{{m{C_R}}}{{{{\bar \gamma }_{{\text{RF}}}}}},\frac{{{\mu _1}}}{{\alpha \beta }}} \right]} \right).
\end{align}

\subsection{DF Relaying}
\subsubsection{Outage Probability} The outage probability of DF relaying can be obtained by using \eqref{DF_End_to_End_CDF} which is given as \cite{8727983,8715395}
\begin{equation}
\label{DF_OP}
P_{_{\text{out}}}^D\left( {{\gamma _{\text{th}}}} \right) = \Pr \left( {{\gamma ^D} < {\gamma _{\text{th}}}} \right) = {F_{{\gamma ^D}}}\left( {{\gamma _{\text{th}}}} \right).
\end{equation}

\subsubsection{Average BER} Substituting \eqref{DF_End_to_End_CDF} into \eqref{AF_AEBR_PRI}, using \cite[Eq.  (2.25.1/1)]{26} along with \cite[Eq. (2.3)]{23} after some algebraic manipulations, the average BER can be obtained in closed-form as
\begin{align}
\label{DF_ABER}
&{{\bar P}_e}^D \!=\! \frac{{n\delta }}{2} \!-\! \frac{{{\xi ^2}{m^m}}}{{\Gamma \!\left( \alpha  \right)\!\Gamma \!\left( \beta  \right)\!\Gamma \!\left( m \right)}}\!\sum\limits_{j = 0}^\infty  \!{\frac{{{K^j}{d_j}}}{{j!\Gamma \left( {j + 1} \right)}}} \!\sum\limits_{k = 1}^n {H_{1,0:1,2:2,4}^{0,1:2,0:4,0}} \nonumber\\
 &\times \left[ {\left. {\begin{array}{*{20}{c}}
{\begin{array}{*{20}{c}}
{\begin{array}{*{20}{c}}
{\begin{array}{*{20}{c}}
{\begin{array}{*{20}{c}}
{\left( {1 - p,1,1} \right)}\\
 -
\end{array}}\\
{\left( {1,1} \right)}
\end{array}}\\
{\left( {0,1} \right)\left( {j + 1,1} \right)}
\end{array}}\\
{\left( {{\xi ^2} + 1,1} \right)\left( {r,1} \right)}
\end{array}}\\
{\left( {0,1} \right)\left( {{\xi ^2},r} \right)\left( {\alpha ,r} \right)\left( {\beta ,r} \right)}
\end{array}} \right|\frac{1}{{2{\sigma ^2}{q_k}}},\frac{{{{\left( {\alpha \beta } \right)}^r}}}{{{\mu _r}{q_k}}}} \right].
\end{align}
Then the asymptotic BER is derived as in \eqref{DF_ABER_ASY} by substituting \eqref{DF_End_to_End_CDF_ASY} into \eqref{AF_AEBR_PRI} and utilizing \cite[Eq. (3.351.3)]{19}.
\begin{align}
\label{DF_ABER_ASY}
&{{\bar P}^D}_e\! \approx\! \frac{\delta }{{2\Gamma\! \left( p \right)}}\!\sum\limits_{k = 1}^n\! {\frac{{r\Gamma\! \left( {\alpha  \!-\! {\xi ^2}} \right)\Gamma \!\left( {\beta  \!-\! {\xi ^2}} \right)\Gamma \!\left( {p \!+ \! \frac{{{\xi ^2}}}{r}} \right)}}{{\Gamma\! \left( \alpha  \right)\Gamma\! \left( \beta  \right)}}} {\left( \!{\frac{{{{\left( {\alpha \beta } \right)}^r}}}{{{\mu _r}{q_k}}}} \!\right)^{\frac{{{\xi ^2}}}{r}}}\nonumber\\
 &+ \frac{\delta }{{2\Gamma \left( p \right)}}\sum\limits_{k = 1}^n {\frac{{r{\xi ^2}\Gamma \left( {{\xi ^2} \!-\! \alpha } \right)\Gamma \left( {\beta  \!-\! \alpha } \right)\Gamma \left( {p \!+\! \frac{\alpha }{r}} \right)}}{{\alpha \Gamma \left( \alpha  \right)\Gamma \left( \beta  \right)\Gamma \left( {{\xi ^2} \!+\! 1 \!-\! \alpha } \right)}}} {\left( {\frac{{{{\left( {\alpha \beta } \right)}^r}}}{{{\mu _r}{q_k}}}} \right)^{\frac{\alpha }{r}}}\nonumber\\
 &+ \frac{\delta }{{2\Gamma \left( p \right)}}\sum\limits_{k = 1}^n {\frac{{r{\xi ^2}\Gamma \left( {{\xi ^2}\! -\! \beta } \right)\Gamma \left( {\alpha  \!-\! \beta } \right)\Gamma \left( {p \!+\! \frac{\beta }{r}} \right)}}{{\beta \Gamma \left( \alpha  \right)\Gamma \left( \beta  \right)\Gamma \left( {{\xi ^2} \!+\! 1 \!-\! \beta } \right)}}} {\left( {\frac{{{{\left( {\alpha \beta } \right)}^r}}}{{{\mu _r}{q_k}}}} \right)^{\frac{\beta }{r}}}\nonumber\\
 &+ \!\frac{\delta }{{2\Gamma \!\left( p \right)}}\!\sum\limits_{k = 1}^n \!{\frac{{{m^m}}}{{\Gamma \!\left( m \right)}}\!\sum\limits_{j = 0}^\infty \! {\frac{{{K^j}{d_j}\Gamma\! \left( {p + j + 1} \right)}}{{j!\left( {j + 1} \right)\Gamma \!\left( {j + 1} \right)}}}\! {{\left(\! {\frac{1}{{2{\sigma ^2}{q_k}}}}\! \right)}^{j + 1}}},
\end{align}
the diversity order of DF relaying is found as
\begin{equation}
\label{DF_Diversity_Gain}
{G_d} = \min \left( {1,\frac{{{\xi ^2}}}{r},\frac{\alpha }{r},\frac{\beta }{r}} \right).
\end{equation}

\subsubsection{Ergodic Capacity}
We can rewrite ${\left( {1 + c\gamma } \right)^{ - 1}}$ in terms of the Fox's $H$-function utilizing \cite[Eq. (1.43)]{24} as $H_{1,1}^{1,1}\left[ {c\gamma \left| {\begin{array}{*{20}{c}}
{\left( {0,1} \right)}\\
{\left( {0,1} \right)}
\end{array}} \right.} \right]$, then substituting \eqref{DF_End_to_End_CDF} into \eqref{AF_EC_PRI}, we can derive the integral of the product of three Fox's $H$-functions. Using \cite[Eq. (2.25.1/1)]{26} and \cite[Eq. (2.3)]{23}, we obtain the ergodic capacity of DF relaying as shown in \eqref{DF_EC} at the bottom of this page. As we can see clearly form \eqref{DF_EC}, the ergodic capacity decreases as the values of $m$ decrease, which is due to the fact that smaller $m$ represents more severe fading in the RF link.
\subsubsection{Effective Capacity}By representing ${{{\left( {1 + \gamma } \right)}^{ - A - 1}}}$ in terms of the Fox's $H$-function with the help of \cite[Eq. (1.43)]{24} as ${H_{1,1}^{1,1}\left[ {\gamma \left| {\begin{array}{*{20}{c}}
{\left( { - A,1} \right)}\\
{\left( {0,1} \right)}
\end{array}} \right.} \right]}$, then substituting \eqref{DF_End_to_End_CDF} into \eqref{AF_EffectiveCapacity_PRI}, using \cite[Eq. (2.25.1/1)]{26} and \cite[Eq. (2.3)]{23}, the effective capacity of DF relaying can be derived as shown in \eqref{DF_Effective} at the bottom of this page.
It should be mentioned that when $\Delta  = 0$, $K \to \infty $, \eqref{DF_Effective} reduces to the ergodic capacity
over mixed FSO/mmWave systems in Gamma-Gamma/Nakagami-$m$ fading channels with heterodyne detection as given by
\begin{align}
%\label{DF_Eff_SpecialCase}
&{R^D} =  - \frac{1}{A}{\log _2}\left( {1 - \frac{{A{\xi ^2}{\mu _1}}}{{\alpha \beta \Gamma \left( \alpha  \right)\Gamma \left( \beta  \right)\Gamma \left( m \right)}}H_{4,2:1,1:1,2}^{0,4:1,1:2,0}} \right.\nonumber\\
&\left. { \times \left[ \!\!\!\!\!\!\!\!\!\!\!\!\!\!\!\!{\left. {\begin{array}{*{20}{c}}
{\begin{array}{*{20}{c}}
{\begin{array}{*{20}{c}}
{\begin{array}{*{20}{c}}
{\begin{array}{*{20}{c}}
{\left( {0;1,1} \right)\left( { - {\xi ^2};1,1} \right)\left( { - \alpha ;1,1} \right)\left( { - \beta ;1,1} \right)}\\
{\left( { - {\xi ^2} - 1;1,1} \right)\left( { - 1;1,1} \right)}
\end{array}}\\
{\left( { - A,1} \right)}
\end{array}}\\
{\left( {0,1} \right)}
\end{array}}\\
{\left( {1,1} \right)}
\end{array}}\\
{\left( {0,1} \right)\left( {m,1} \right)}
\end{array}} \!\!\!\!\!\!\!\!\!\!\!\!\!\right|\frac{{{\mu _1}}}{{\alpha \beta }},\frac{{m{\mu _1}}}{{\alpha \beta {{\bar \gamma }_{{\text{RF}}}}}}} \right]} \right).\notag
\end{align}

\section{Numerical Results} In this section, we compare the analytical results against Monte Carlo simulations to verify our derived expressions. We assume equal average SNRs of both the links, i.e., ${\mu _r} = {{\bar \gamma }_{\text{RF}}} = \bar \gamma $. Specifically,  for the FSO link, two different channel parameters $\left( {\alpha ,\beta } \right) = \left( {{\rm{5}}{\rm{.42}},3.8} \right)$ and $\left( {\alpha ,\beta } \right) = \left( {3.446,1.032} \right)$ are considered to represent moderate and strong turbulence conditions, respectively. $\xi  = 0.893$ and $\xi  = 5.0263$ are used to represent strong and negligible pointing errors, respectively. For the RF link, we set the fading figure as $m = 0.3$ and $m = 2$. Furthermore, a fixed relay gain ${C_R} = 1.7$ is considered.

Figure 2 shows the impact of strong $\left( {\xi  = 0.893} \right)$ and negligible $\left( {\xi  = 5.0263} \right)$ pointing errors on the outage probability performance of dual-hop mixed FSO/mmWave systems using fixed-gain AF relaying under strong and moderate turbulence conditions, with the RF link parameters $K = 10$, $m = 2$ and $\Delta  = 0.5$. It can be observed from this figure that the analytical results match perfectly with the MATLAB simulated results and the accuracy of our derivation is proved. Moreover, as expected, the higher the values of $\xi $ , the lower will be the outage probability. Furthermore, strong turbulence conditions lead to higher outage probability compared with moderate turbulence. We can also observe that the asymptotic expression derived in \eqref{AF_End_to_End_CDF_ASY} gives tight asymptotic results in the high-SNR regime. In addition, we have ${G_d} = \frac{{{\xi ^2}}}{r} < 2$ under strong pointing error, we can further find that ${G_d} = 2$  and ${G_d} = \frac{\beta }{r} < 2$ under negligible pointing error for moderate turbulence and strong turbulence, respectively.

\begin{figure}[t]
\centering
\includegraphics[scale=0.8]{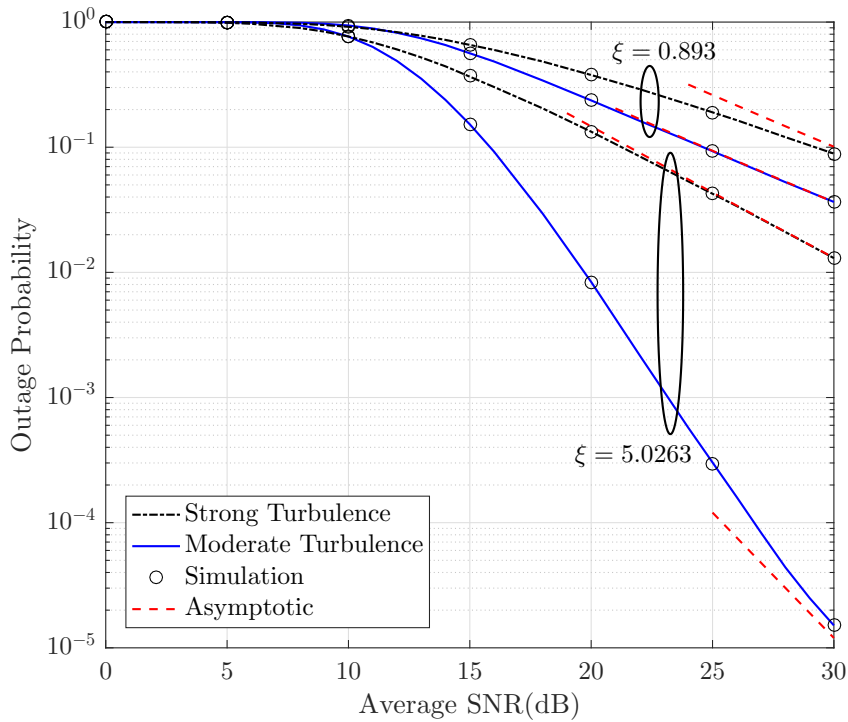}
\caption{Outage probability of a dual-hop mixed FSO/mmWave fixed-gain AF relay system for
strong and negligible pointing errors under different turbulence conditions
using heterodyne detection ($K = 10$, $m = 2$ and $\Delta  = 0.5$).
\label{OPfigure}}
\end{figure}

\begin{figure}[t]
\centering
\includegraphics[scale=0.8]{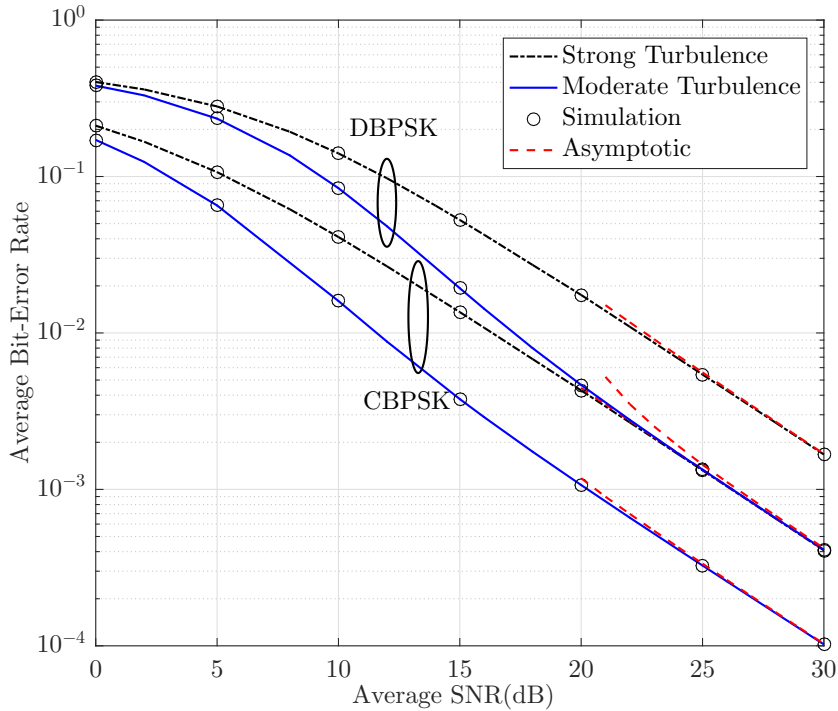}
\caption{Average BER of a dual-hop mixed FSO/mmWave system using DF relaying
for strong and moderate turbulence conditions under heterodyne detection ($\xi =5.0263$, $K = 10$, $m = 2$, and $\Delta  = 0.5$).
\label{DF_ABRE1figure}}
\end{figure}

\begin{figure}[t]
\centering
\includegraphics[scale=0.8]{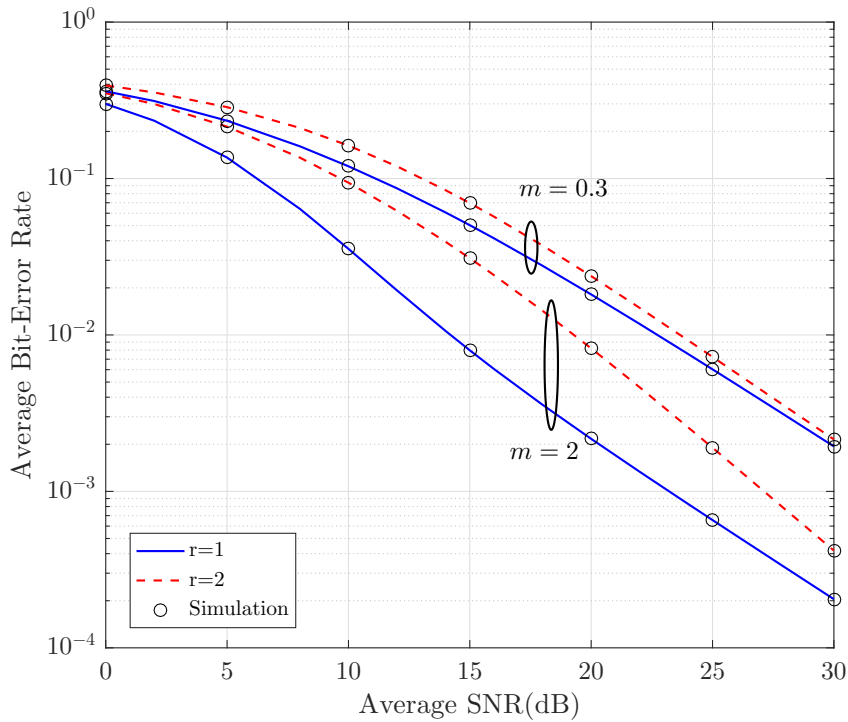}
\caption{Average BER of a dual-hop mixed FSO/mmWave DF system using DBPSK,
 IM/DD and heterodyne techniques with negligible pointing errors ($K = 10$ and $\Delta  = 0.5$).
\label{DF_ABRE2figure}}
\end{figure}

\begin{figure}[t]
\centering
\includegraphics[scale=0.8]{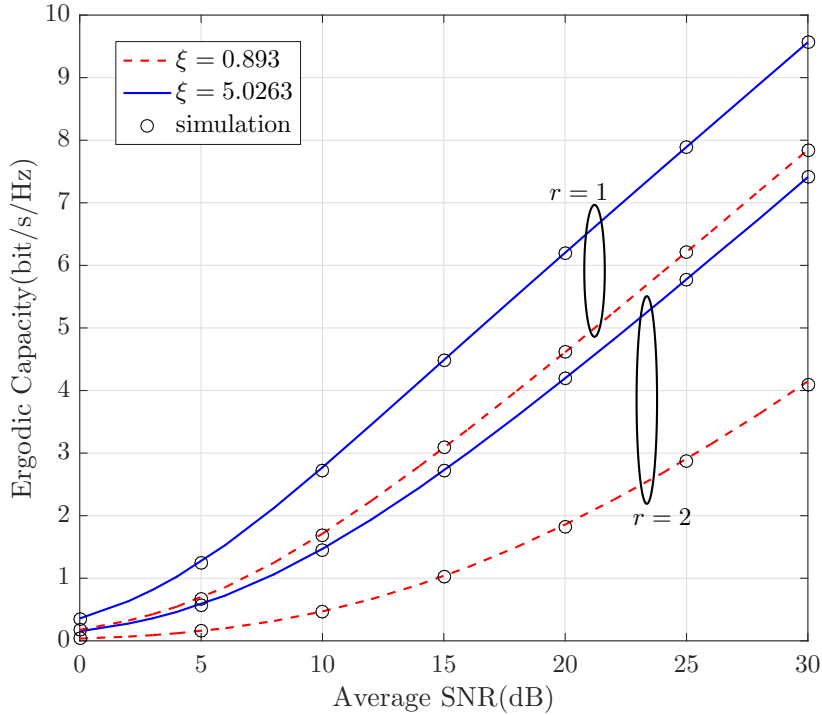}
\caption{Ergodic capacity of a dual-hop mixed FSO/mmWave system using fixed-gain AF relaying
for strong and negligible pointing errors  ($K = 10$, $m = 2$ and $\Delta  = 0.5$).
\label{AF_ECfigure}}
\end{figure}

\begin{figure}[t]
\centering
\includegraphics[scale=0.8]{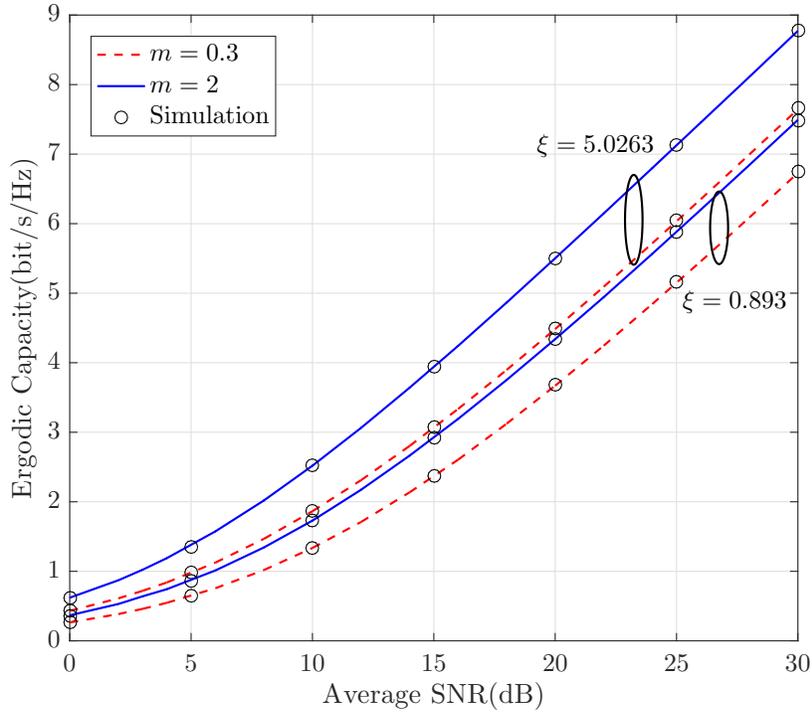}
\caption{Ergodic capacity of DF relaying for moderate
turbulence under heterodyne detection ($K = 10$ and $\Delta  = 0.5$).
\label{DF_ECfigure}}
\end{figure}

\begin{figure}[t]
\centering
\includegraphics[scale=0.8]{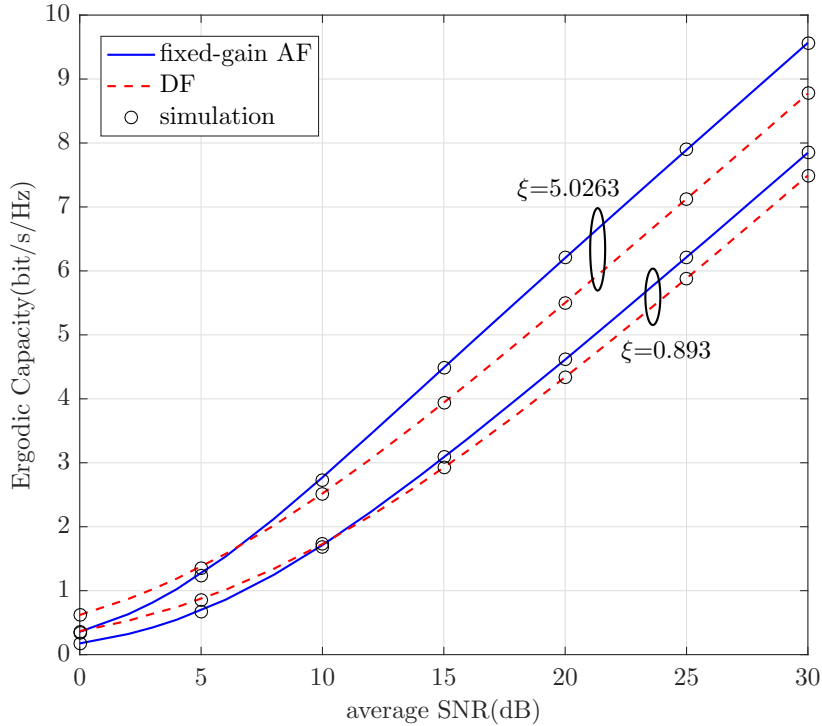}
\caption{Ergodic capacity comparison of different relaying protocols for
strong and negligible pointing errors under moderate turbulence and heterodyne technique($K = 10$, $m = 2$ and $\Delta  = 0.5$).
\label{AF&DF}}
\end{figure}

\begin{figure}[t]
\centering
\includegraphics[scale=0.8]{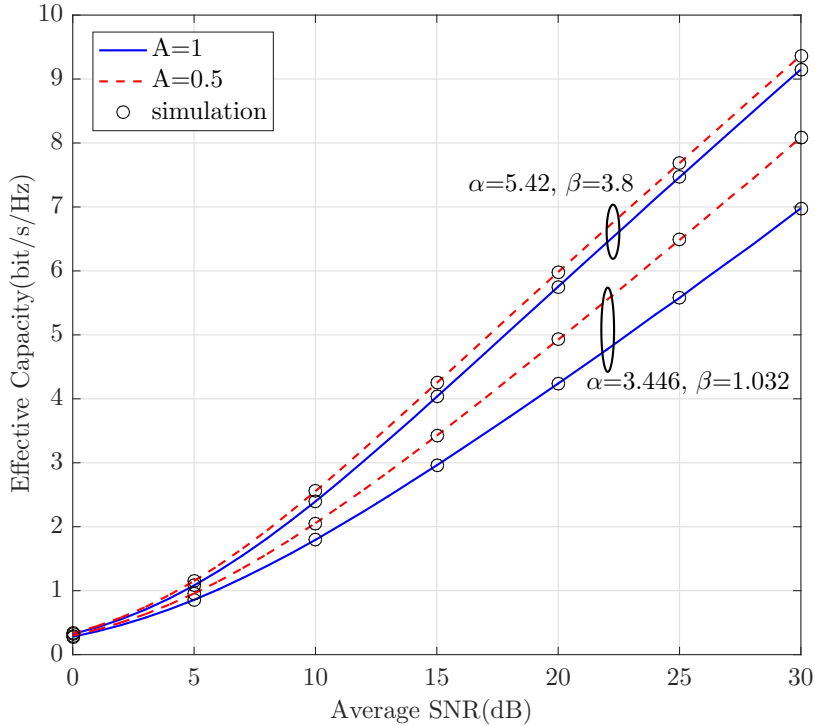}
\caption{Effective capacity of a dual-hop mixed FSO/mmWave system using
fixed-gain AF relaying with strong and moderate turbulence
 and different values of A. ($\xi =5.0263$, $K = 2$, $m = 2$ and $\Delta  = 0.5$).
\label{AF_Effective}}
\end{figure}

\begin{figure}[t]
\centering
\includegraphics[scale=0.8]{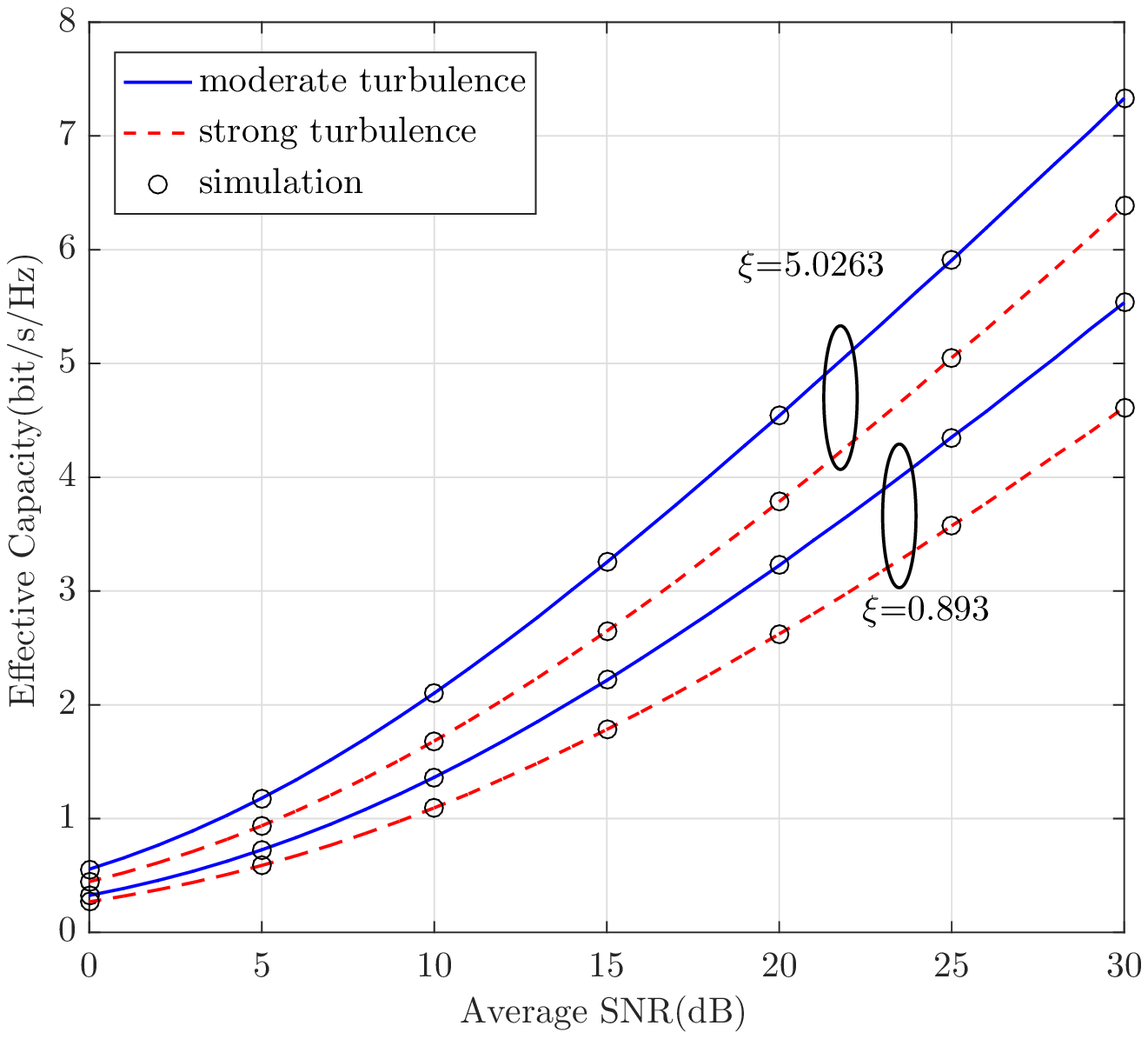}
\caption{Effective capacity of a dual-hop mixed FSO/mmWave system using DF relaying for strong and negligible pointing errors under different turbulence conditions ($A = 1$, $K = 2$, $m = 2$ and $\Delta  = 0.5$).
\label{DF_Effectivefig}}
\end{figure}

Figure 3 illustrates the average BER performance in the case of moderate and strong turbulence regimes for DBPSK and CBPSK binary modulation schemes with fixed effect of the pointing error $\left( {\xi {\rm{ = }}5.0263} \right)$. As expected, the average BER decreases as the average SNR increases under different turbulence conditions for both types of modulation schemes. It can also be observed that the average BER performance is degraded under severe turbulence conditions.

In Fig. 4, the average BER of dual-hop mixed FSO/mmWave systems under both heterodyne and IM/DD techniques for DBPSK modulation scheme is presented over different fading severity parameters, $m$. It can be observed that IM/DD behaves worse than heterodyne detection, which can be explained by the fact that heterodyne technique can better overcome the turbulence effects. Moreover, it also shows severe fading in the RF link $\left( {m = 0.3} \right)$ increases the average BER for both types of detection techniques. We can also observe that the effect of the fading parameter on the system performance is more significant when heterodyne technique is employed compared to IM/DD case.

In Fig. 5, we present the ergodic capacity when fixed-gain AF relaying is employed in operation under both heterodyne detection and IM/DD techniques for strong and negligible pointing errors with moderate turbulence. Expectedly, we can see from this figure that the heterodyne detection performs better than IM/DD. Furthermore, we can also observe that the stronger the effect of pointing error, the lower is the ergodic capacity of the system.

We compare the ergodic capacity using DF relaying under heterodyne technique for different fading figures assuming moderate turbulence conditions in Fig. 6. It is observed that by increasing the average SNR, the ergodic capacity performance is improved. In addition, it can be seen that the ergodic capacity decreases as $m$ decreases, which represents that the effect of the fading in the RF link gets severe. It also shows that negligible pointing error yields better performance in ergodic capacity compared with strong pointing error.

Figure 7 investigates the ergodic capacity performance of mixed FSO/mmWave systems assuming both AF and DF relaying for strong and negligible pointing errors over moderate turbulence and heterodyne technique conditions. It is observed that when the average SNR is low, DF relaying performs better than fixed-gain AF relaying in ergodic capacity. However, as the average SNR increases, fixed-gain AF relaying yields higher ergodic capacity. Furthermore, it can be inferred from Fig. 7
that as the pointing error gets severe, the ergodic capacity decreases.

Figure 8 depicts the effective capacity when fixed-gain AF relaying and heterodyne detection technique is employed in operation with negligible pointing errors and different values of A under the effects of the moderate and strong turbulence conditions. By observing the results in Fig. 8, it is clear that the more severe is the atmospheric turbulence, the higher is the degradation in the effective capacity performance under both values of the normalized QoS exponent $A$. In addition, Fig. 8 shows that the effective capacity increases when $A$ becomes smaller, which can be explained that $A$ represents the normalized QoS exponent. As the value of $A$ gets smaller, the QoS requirement becomes less stringent.

Figure 9 demonstrates the effective capacity performance comparison results of the dual-hop mixed FSO/mmWave system under DF relaying and heterodyne detection technique to show the impact of different turbulence and pointing error conditions. It can be shown that the effective capacity starts dropping as the turbulence gets worse. Moreover, we can deduce from Fig. 9 that the large values of pointing errors result in significant effective capacity performance degradation. Finally, it is worth mentioned that the analytical and simulation results are again in excellent agreement, thus verifying the correctness of the mathematical analysis previously presented in Section IV.

\section{Conclusions}
In this paper, we have presented novel closed-form performance metrics for dual-hop mixed FSO/mmWave systems using both heterodyne detection and direct detection techniques where the FSO link and the mmWave link experience Gamma-Gamma atmospheric turbulence with pointing errors taken into account and FTR fading, respectively. Assuming both AF and DF relaying, we derived exact closed-form expressions for the outage probability,  average bit error probability, ergodic capacity and effective capacity of the considered system. In addition, we analyzed the asymptotical performance at high SNR regime and derived diversity orders. Our asymptotic results show that diversity orders are related to FSO turbulence parameters, pointing error, and detection method. We further compared the system performance under different turbulence conditions, pointing errors, relaying techniques and fading figures of the mmWave RF link. As expected, weak turbulence conditions, negligible pointing errors and AF relaying with an increase of the fading figure can improve the system performance.

\begin{appendices}
\section{Proof of Corollary \ref{coro:CDF}}
We can derive the CDF of the end-to-end SNR ${\gamma ^F}$ as
\setcounter{equation}{0}
\renewcommand{\theequation}{\thesection.\arabic{equation}}
\begin{align}
\label{EA.1}
&{F_{{\gamma ^F}}}\left( \gamma  \right)= {\rm{ Pr}}\left( {\frac{{{\gamma _{\text{FSO}}}{\gamma _{\text{RF}}}}}{{{\gamma _{\text{RF}}} + {C_R}}} < \gamma } \right)\nonumber\\
 %&= \int_0^\infty  {\Pr \left[ {\frac{{x{\gamma _{\text{RF}}}}}{{{\gamma _{\text{RF}}} + {C_R}}} < \gamma \left| x \right.} \right]} {f_{{\gamma _{\text{FSO}}}}}\left( x \right)dx\nonumber\\
% &=\int_0^\gamma  {\Pr \left[ {{\gamma _{\text{RF}}} < \frac{{{C_R}\gamma }}{{x - \gamma }}\left| x \right.} \right]} {f_{{\gamma _{\text{FSO}}}}}\left( x \right)dx\nonumber\\
% &+ \int_\gamma ^\infty  {\Pr \left[ {{\gamma _{\text{RF}}} < \frac{{{C_R}\gamma }}{{x - \gamma }}\left| x \right.} \right]} {f_{{\gamma _{\text{FSO}}}}}\left( x \right)dx\nonumber\\
% &= {F_{{\gamma _{\text{FSO}}}}}\left( \gamma  \right) + \int_\gamma ^\infty  {{F_{{\gamma _{\text{RF}}}}}\left( {\frac{{{C_R}\gamma }}{{t - \gamma }}} \right)} {f_{{\gamma _{\text{FSO}}}}}\left( t \right)dt\nonumber\\
 &= 1 - \int_0^\infty  {{f_{{\gamma _{\text{FSO}}}}}\left( {x + \gamma } \right)} \left[ {1 - {F_{{\gamma _{\text{RF}}}}}\left( {\frac{{{C_R}\gamma }}{x}} \right)} \right]dx,
\end{align}
By substituting \eqref{FSO_PDF}, \eqref{RF_CDF} into \eqref{EA.1} and using the definition of the Meijer's $G$-function \cite[Eq. (9.301)]{19}, we write \eqref{EA.1}   as
\begin{align}
\label{EA.2}
&{F_{{\gamma ^F}}}\left( \gamma  \right) = 1 - \frac{{{\xi ^2}{m^m}}}{{r\Gamma \left( \alpha  \right)\Gamma \left( \beta  \right)\Gamma \left( m \right)}}\sum\limits_{j = 0}^\infty  {\frac{{{K^j}{d_j}}}{{j!\Gamma \left( {j + 1} \right)}}} {\left( {\frac{1}{{2\pi i}}} \right)^2}\nonumber\\\
 &\times\! \int\limits_{{C_1}} \!{\int\limits_{{C_2}} \!{\Gamma\! \left( {{s_1} \!-\! \frac{{{s_2}}}{r}} \right)\!\frac{{\Gamma \!\left( {1 \!-\! {s_1}} \right)\Gamma \!\left( { - {s_1}} \right)\Gamma \!\left( {j \!+\! 1 \!-\! {s_1}} \right)}}{{\Gamma \!\left( {1 \!-\! {s_1}} \right)}}{{\left( {\frac{{{C_R}}}{{2{\sigma ^2}}}} \right)}^{{s_1}}}} } \nonumber\\\
 &\times\! \frac{{\Gamma \!\left( {{\xi ^2} - {s_2}} \right)\!\Gamma \!\left( {\alpha  - {s_2}} \right)\!\Gamma \!\left( {\beta  - {s_2}} \right)}}{{\Gamma\! \left( {{\xi ^2} + 1 - {s_2}} \right)\Gamma \!\left( {1 - \frac{{{s_2}}}{r}} \right)}}{\left( \!{\alpha \beta {{\left( {\frac{\gamma }{{{\mu _r}}}} \right)}^{\frac{1}{r}}}} \!\right)^{{s_2}}}d{s_1}d{s_2},
\end{align}
where ${\mathcal{C}_1}$ and ${\mathcal{C}_2}$ represent the ${s_1}$-plane and the ${s_2}$-plane contours, respectively. By utilizing \cite[Eq. (3.194/3)]{19} and \cite[Eq. (8.384/1)]{19}, $\int_0^\infty  {\frac{{{{\left( {x + \gamma } \right)}^{\frac{{{s_2}}}{r}}}}}{{x + \gamma }}{x^{ - {s_1}}}} dx$ simplifies to $\frac{{\Gamma \left( {1 - {s_1}} \right)\Gamma \left( {{s_1} - \frac{{{s_2}}}{r}} \right)}}{{\Gamma \left( {1 - \frac{{{s_2}}}{r}} \right)}}{\gamma ^{\frac{{{s_2}}}{\gamma } - {s_1}}}$ and \eqref{EA.2} becomes
\begin{align}
\label{EA.3}
&{F_{{\gamma ^F}}}\left( \gamma  \right) = 1 - \frac{{{\xi ^2}}}{{r\Gamma \left( \alpha  \right)\Gamma \left( \beta  \right)}}\frac{{{m^m}}}{{\Gamma \left( m \right)}}\sum\limits_{j = 0}^\infty  {\frac{{{K^j}{d_j}}}{{j!\Gamma \left( {j + 1} \right)}}} {\left( {\frac{1}{{2\pi i}}} \right)^2}\nonumber\\
 &\times\! \int\limits_{{C_1}}\! {\int\limits_{{C_2}} {\Gamma \!\left( {{s_1} \!-\! \frac{{{s_2}}}{r}} \right)\!\frac{{\Gamma \!\left( {1 \!-\! {s_1}} \right)\Gamma \!\left( { - {s_1}} \right)\Gamma \!\left( {j \!+\! 1 \!-\! {s_1}} \right)}}{{\Gamma\! \left( {1 \!-\! {s_1}} \right)}}{{\left( {\frac{{{C_R}}}{{2{\sigma ^2}}}} \right)}^{{s_1}}}} } \nonumber\\
 &\times \frac{{\Gamma \!\left( {{\xi ^2} - {s_2}} \right)\Gamma\! \left( {\alpha  - {s_2}} \right)\Gamma \!\left( {\beta  - {s_2}} \right)}}{{\Gamma\! \left( {{\xi ^2} + 1 - {s_2}} \right)\Gamma\! \left( {1 - \frac{{{s_2}}}{r}} \right)}}{\left(\! {\alpha \beta {{\left( {\frac{\gamma }{{{\mu _r}}}} \right)}^{\frac{1}{r}}}} \!\right)^{{s_2}}}d{s_1}d{s_2}.
\end{align}
Finally, by changing the integral variable ${s_2} \to  - {s_2}$, utilizing \cite[Eq. (1.1)]{23} and after some simple algebraic manipulations, we can obtain \eqref{AF_End_to_End_CDF} to finish the proof.

\section{Proof of Corollary \ref{coro:AEBR}}
\setcounter{equation}{0}
\renewcommand{\theequation}{\thesection.\arabic{equation}}
Substituting \eqref{EA.3} into \eqref{AF_AEBR_PRI}, the average BER is given by
\begin{align}
\label{EB.1}
&\bar P_e^F = \frac{{n\delta }}{2} - \frac{\delta }{{2\Gamma \left( p \right)}}\sum\limits_{k = 1}^n {q_k^p\frac{{{\xi ^2}}}{{r\Gamma \left( \alpha  \right)\Gamma \left( \beta  \right)}}\frac{{{m^m}}}{{\Gamma \left( m \right)}}} {\left( {\frac{1}{{2\pi i}}} \right)^2}\nonumber\\
 &\times \sum\limits_{j = 0}^\infty  {\frac{{{K^j}{d_j}}}{{j!\Gamma \left( {j + 1} \right)}}} \int\limits_{{C_1}} {\int\limits_{{C_2}} {\Gamma \left( {{s_1} + \frac{{{s_2}}}{r}} \right)\Gamma \left( { - {s_1}} \right)\Gamma \left( {j + 1 - {s_1}} \right)} } \nonumber\\
 &\times {\left( \!{\frac{{{C_R}}}{{2{\sigma ^2}}}} \!\right)^{{s_1}}}\!\frac{{\Gamma \!\left( {{\xi ^2} + {s_2}} \right)\!\Gamma\! \left( {\alpha  + {s_2}} \right)\!\Gamma \!\left( {\beta  + {s_2}} \right)}}{{\Gamma\! \left( {{\xi ^2} + 1 + {s_2}} \right)\!\Gamma \!\left( {1 + \frac{{{s_2}}}{r}} \right)}}{\left( \!{\frac{{\mu _r^{{1 \mathord{\left/
 {\vphantom {1 r}} \right.
 \kern-\nulldelimiterspace} r}}}}{{\alpha \beta }}} \!\right)^{{s_2}}}\!d{s_1}d{s_2}\nonumber\\
 &\times \int_0^\infty  {{\gamma ^{p - 1}}{\gamma ^{ - {{{s_2}} \mathord{\left/
 {\vphantom {{{s_2}} r}} \right.
 \kern-\nulldelimiterspace} r}}}\exp \left( { - {q_k}\gamma } \right)d\gamma }.
\end{align}
Using \cite[Eq. (3.381/4)]{19} along with \cite[Eq. (1.1)]{23}, we can obtain \eqref{AF_AEBR} to finish the proof.

\section{Proof of Corollary \ref{coro:capa}}
\setcounter{equation}{0}
\renewcommand{\theequation}{\thesection.\arabic{equation}}
Substituting \eqref{AF_End_to_End_CDF} into \eqref{AF_EC_PRI}, we have the ergodic capacity as
\begin{align}
%\label{EC.1}
&{{\bar C}^F} = \frac{{c{\xi ^2}{m^m}}}{{\ln \left( 2 \right)r\Gamma \left( \alpha  \right)\Gamma \left( \beta  \right)\Gamma \left( m \right)}}\sum\limits_{j = 0}^\infty  {\frac{{{K^j}{d_j}}}{{j!\Gamma \left( {j + 1} \right)}}} \frac{1}{{{{\left( {2\pi i} \right)}^2}}}\nonumber\\
& \!\times\!\! \int\limits_{{C_1}}\! {\int\limits_{{C_2}}\! {\frac{{\Gamma \!\left( {{s_1} \!+\! \frac{{{s_2}}}{r}} \right)\Gamma \!\left( {{\xi ^2} \!+\! {s_1}} \right)\Gamma\! \left( {\alpha  \!+\! {s_1}} \right)\Gamma \!\left( {\beta  \!+\! {s_1}} \right)}}{{\Gamma \!\left( {1 \!+\! {\xi ^2} + {s_1}} \right)\Gamma\! \left( {1 \!+ \frac{{{s_1}}}{r}} \right)}}{{\left(\! {\frac{1}{{\alpha \beta }}{\mu _r}^{\frac{1}{r}}}\! \right)}^{{s_1}}}} } \nonumber\\
 &\!\times \Gamma \left( { - {s_2}} \right)\Gamma \left( {j + 1 - {s_2}} \right){\left( {\frac{{{C_R}}}{{2{\sigma ^2}}}} \right)^{{s_2}}}d{s_1}d{s_2}\int_0^\infty  {\frac{{{\gamma ^{ - \frac{{{s_1}}}{r}}}}}{{1 + c\gamma }}} d\gamma. \notag
\end{align}
Using \cite[Eq. (4.293/10)]{19} and \cite[Eq. (1.1)]{23}, the proof is finished by deriving \eqref{AF_EC}.

\section{Proof of Corollary \ref{coro:effective}}
\setcounter{equation}{0}
\renewcommand{\theequation}{\thesection.\arabic{equation}}
Substituting \eqref{AF_End_to_End_CDF} into \eqref{AF_EffectiveCapacity_PRI}, we have the ergodic capacity as
\begin{align}
&{R^F}{\rm{ = }} -\! \frac{1}{A}{\log _2}\!\left(\! {1 - \frac{{A{\xi ^2}{m^m}}}{{r\Gamma \left( \alpha  \right)\Gamma \left( \beta  \right)\Gamma \left( m \right)}}\sum\limits_{j = 0}^\infty  {\frac{{{K^j}{d_j}}}{{j!\Gamma \left( {j + 1} \right)}}} \frac{1}{{{{\left( {2\pi i} \right)}^2}}}} \right.\nonumber\\
 &\times \!\!\int\limits_{{C_1}}\! {\int\limits_{{C_2}} \!{\frac{{\Gamma\! \left( {{s_1} + \frac{{{s_2}}}{r}} \right)\Gamma \!\left( {{\xi ^2} + {s_1}} \right)\Gamma\! \left( {\alpha  + {s_1}} \right)\Gamma\! \left( {\beta  + {s_1}} \right)}}{{\Gamma\! \left( {{\xi ^2} + 1 + {s_1}} \right)\Gamma\! \left( {1 + \frac{{{s_1}}}{r}} \right)}}{{\left(\! {\frac{{{{\left( {{\mu _r}} \right)}^{\frac{1}{r}}}}}{{\alpha \beta }}} \!\right)}^{{s_1}}}} }\nonumber \\
&\left. { \times \Gamma\! \left( { - {s_2}} \right)\!\Gamma\! \left( {j + 1 - {s_2}} \right){{\left( {\frac{{{C_R}}}{{2{\sigma ^2}}}} \right)}^{{s_2}}}d{s_1}d{s_2}\!\!\int_0^\infty\!  {\frac{{{\gamma ^{ - \frac{{{s_1}}}{r}}}}}{{{{\left( {1 + \gamma } \right)}^{A + 1}}}}} d\gamma } \!\right).\notag
\end{align}
Using \cite[Eq. (4.293/10)]{19} and \cite[Eq. (1.1)]{23}, the proof is finished by deriving \eqref{AF_EffectiveCapacity}.

\end{appendices}

%--------------------------------------------------------

\bibliographystyle{IEEEtran}
\bibliography{IEEEabrv,Ref}

\end{document}